\documentclass{ws-ijmpa}
\usepackage{graphicx}
\usepackage{epsfig}
\usepackage{bm}
\usepackage{amsfonts}
\usepackage{epstopdf}
\usepackage{multirow}
\usepackage[colorlinks,linkcolor=red,anchorcolor=blue,citecolor=green]{hyperref}
\usepackage[super,compress]{cite}
\usepackage{subfigure}
\renewcommand{\theequation}{\thesection.\arabic{equation}}
\newcommand\diff{\mathrm{d}}
\newcommand\Diff{\mathrm{D}}
\newcommand\im{\mathrm{i}}
\newcommand\e{\mathrm{e}}
\begin{document}

\title{Some Developments of the Casimir Effect in $p$-Cavity of $(D+1)$-Dimensional Spacetime}

\author{Xiang-Hua Zhai}
\author{Rui-Hui Lin}
\author{Chao-Jun Feng}
\author{Xin-Zhou Li}

\address{Shanghai United Center for Astrophysics (SUCA), Shanghai Normal University,\\
    100 Guilin Road, Shanghai 200234, China\\
    zhaixh@shnu.edu.cn
    }

\maketitle

\begin{abstract}
The Casimir effect for rectangular boxes has been studied for several decades.
But there are still some points unclear.
Recently, there are new developments related to this topic,
including the demonstration of the equivalence of the regularization methods
and the clarification of the ambiguity in the regularization of the temperature-dependent free energy.
Also, the interesting quantum spring was raised
stemming from the topological Casimir effect of the helix boundary conditions.
We review these developments together with the general derivation of the Casimir energy of the $p$-dimensional cavity
in ($D+1$)-dimensional spacetime, paying special attention to the sign of the Casimir force in a cavity with unequal edges.
In addition, we also review the Casimir piston, which is a configuration related to rectangular cavity.

\end{abstract}
\ccode{PACS numbers: 03.70.+k, 11.10.-z}

\section{Introduction}
\label{intro}
The Casimir effect, as the embodiment of the quantum fluctuation,
since its first prediction\cite{Casimir1948} more than 60 years ago,
has been put under extensive and detailed study theoretically and experimentally.
Yet it still receives increasing attention from the scientific community.
The nature of this effect, many aspects of which have been reviewed in a large amount of literature
\cite{Plunien1986,Bordag2009,Bordag2001,milonni1994,milton2001casimir,mostepanenko1997},
may depend on the background field, the geometry of the configuration,
the type of boundary conditions (BCs), the topology of spacetime, the spacetime dimensionality,
and the finite temperature.
As a simple generalization of the original setup of the two parallel planes\cite{Casimir1948},
the Casimir effect in rectangular boxes,
is one of the frequently considered configuration,
and has been a topic for several decades.
Various calculation methods have been developed for this configuration.
(For an example see Ref. \refcite{Bordag2009} and references therein
and also Refs. \citen{Fierz1960,Boyer1970,Lukosz1971,Mamaev1979,Ambjorn1983,Mamaev1979a,Verschelde1984,Svaiter1991,Caruso1991,Hacyan1993,Actor1994,Gilkey1994,Actor1995,Actor1995a,Li1997,mostepanenko1997,Zheng1998,Maclay2000,Santos2000,Barton2001,Li2001,Edery2003,Graham2003,Inui2003,Hertzberg2005,Alves2006,Barton2006,Gusso2006,Jauregui2006,Lim2007,Edery2007,Hertzberg2007,Zhai2007,ZHAI2009,Rypestol2010,Teo2010,FENG2014,Lin2014,Lin2014a}).

In the calculation and regularization of Casimir effect inside a rectangular box,
the commonly used methods are Abel-Plana formula and zeta function technique\cite{Dowker1976,Hawking1977,Kirsten1991,ELIZALDE1992,Elizalde1994a,Elizalde1995,Cougo-Pinto1999,Elizalde2008,Elizalde2012,Elizalde2013,Erdas2013}.
The zeta function technique\cite{Dowker1976,Hawking1977,Elizalde1994a,Elizalde1995,Elizalde2008,Elizalde2012},
which can be traced back to G. H. Hardy\cite{Hardy1916,Hardy1949},
is used to be regarded as an elegant and unique regularization method\cite{Elizalde1994,Ortenzi2004}
different from other ones such as frequency cut-off method (see e.g. Ref. \refcite{Boyer1968})
and Abel-Plana formula (see e.g. Refs. \refcite{Barton1981,Barton1982}).
Although the automatically finite result of this method veils the isolation of the divergent part,
the fact that in most cases the outcomes are in agreement with other approaches
\cite{Ruggiero1977,Ruggiero1980,Blau1988,Elizalde1994a,Bordag2001,Cavalcanti2004,Edery2006,Frank2008,Herdeiro2008,Bellucci2009,Bellucci2009a,Bordag2009,Linares2010}
draws some attention to the investigation of zeta function itself\cite{Kirsten2001,Tierz2001,Bordag2001,Bordag2009}
and its connection with other regularization methods\cite{Reuter1985,Svaiter1991,Svaiter1992,Beneventano1996,Kurokawa2002}.
As a matter of fact,
the divergent part of zeta function which is implicitly removed
is shown and regulated for two and three-dimensional boxes\cite{Bordag2009}
by utilization and comparison of the Abel-Plana formula method,
which permits explicit separation of the infinite terms.
And now, the two methods are proved to be identifiable\cite{Lin2014b},
that is, the reflection formula of Epstein zeta function,
which is the key of the regularization process,
can be derived from the Abel-plana formula.
So, one can choose any methods for convenience.
We shall come back to the demonstration of the equivalence of these two regularization methods in Sect. \ref{equivreg}.

With the help of the powerful and facile technique of zeta function,
the dependence of the Casimir effect on the configurations of the box is within reach.
Specifically, the attractive or repulsive nature of the force depending on the configuration
is a subject of concern in the study of Casimir effect in rectangular boxes\cite{Lukosz1971,Boyer1974,Mamaev1979,Mamaev1979a,Unwin1982,Ambjorn1983,Caruso1991,Hacyan1993,Li1997,Cougo-Pinto1999,Maclay2000,Barton2001,Li2001,Kenneth2002,Edery2003,Iannuzzi2003,Pinto2003,Bordag2009},
and is analysed much conveniently in terms of zeta function.
We will give a general derivation of the Casimir energy under various BCs
and discuss the result for both equal and unequal edges in Sect. \ref{pmbox},
where the previous results in the literature are recovered as special cases.
Especially, we review the repulsive force caused by unequal edges under Dirichlet BCs\cite{Li1997}.

The calculations in the rectangular box indicate that
the Casimir energy may change sign depending not only on the BCs
but also on geometry of the configuration.
To address the doubt of the repulsive force,
the configuration of a rectangular piston, a box divided by an ideal movable partition,
is proposed\cite{Cavalcanti2004}.
For a scalar field obeying Dirichlet BCs on all surfaces,
when the separation between the piston and one end of the cavity approaches infinity,
the force on the piston is towards another end (the closed end),
that is, the force is always attractive, independent of the ratio of the edges\cite{Zhai2007,ZHAI2009}.
Now it is known that the results are in agreement because the two configurations are actually different\cite{Mostepanenko2009}.
And then, the Casimir effect on various piston geometries and for various fields under various BCs,
and also with various spacetime dimensions attracts a lot of interests
\cite{Kenneth2002,Iannuzzi2003,Hertzberg2005,Barton2006,Edery2007,Rodriguez2007,Edery2007a,Fulling2007,Marachevsky2007,Zhai2007,Cheng2008,Marachevsky2008,Edery2008,Edery2008a,Schaden2008,ZHAI2009,Lim2009,Lim2009a,OIKONOMOU2009,Teo2009a,Teo2009b,Teo2009c,Lim2009b,Teo2009d,Teo2009e,Edery2009,Kirsten2009,Elizalde2009,Mostepanenko2009,Cheng2010,Cheng2010a,Teo2010a,OIKONOMOU2010,McCauley2010,Teo2010b,Khoo2011,Dowker2011,Fucci2011,Fucci2011a,Teo2011,Oikonomou2011,FUCCI2012,Beauregard2013}.
Both attractive and repulsive forces are obtained under corresponding conditions.
We will focus on the rectangular Casimir piston model of massless and massive scalar field
under Dirichlet and hybrid BCs in Sect. \ref{piston}.

The researches mentioned above are limited to the vacuum state of the quantum field,
namely all the excitation are neglected and the temperature of the system is set to be zero,
which seems not practical nor feasible.
The quantum state containing particles in thermal equilibrium with a finite characteristic temperature $T$
is a typical situation when considering the influence of temperature on the Casimir effect.
Indeed, thermal corrections on the Casimir effect for various configuration did attract a lot of interest
\cite{Lifshitz1956,Dzyaloshinskii1961,Mehra1967,Brown1969,Schwinger1978,Altaie1978,Ambjorn1983,Plunien1987,Kirsten1991,ELIZALDE1992,Santos2000,Cheng2002,Inui2002,Nesterenko2004,Jauregui2006,Lim2007,Hertzberg2007,MOSTEPANENKO2010}.
Both controversies and progesses were seen in this topic\cite{Milton2004,Brevik2006,Lim2007,Geyer2008,Milton2009,Teo2009,Cheng2010,Brevik2013,Erdas2013},
and it is exciting that there is a possibility to measure the thermal effect in the Casimir force
\cite{Geyer2010,KLIMCHITSKAYA2012,KLIMCHITSKAYA2011}.
The Casimir effect at finite temperature for a $p$-dimensional rectangular cavity inside a $(D+1)$-dimensional spacetime
was first considered by Ambj{\o}rn and Wolfram\cite{Ambjorn1983},
and was reconsidered by Lim and Teo\cite{Lim2007} more recently in detail,
expanding the results of different BCs in the low and high temperature regimes.
And critical discussion was given by Geyer et al. on the thermal Casimir effect
in ideal metal rectangular boxes in three-dimensional space\cite{Geyer2008},
pointing out the neglect of the removal of geometrical contributions
including the blackbody radiation term in the previous researches,
which would lead to the contradiction with the classical limit.
Now a common recognition was reached that the terms of order equal to or more than the square of the temperature
should be subtracted from the Casimir energy.
But the explicit expression of these terms is not easy to get
from the calculation of the heat kernel coefficients.
Recently, these terms were obtained\cite{Lin2014,Lin2014a}
by repeatedly using Abel-Plana formula,
and more importantly,
the subtraction of them was shown clearly by rigorous calculation and regularization
of the temperature-dependent part of the thermal scalar Casimir energy and force with different BCs.
We will give the review of these results in Sect. \ref{temperature}.

As mentioned before, besides the geometry, the BCs and the temperature,
the nontrivial topology of the space can also give rise to the Casimir effect.
The scalar field on a flat manifold with topology of a circle $S^1$ and a M\"obius strip may be the simplest examples.
Periodic condition $\phi(t,0)=\phi(t,C)$ caused by the topology of $S^1$ with circumference of $C$,
and similar antiperiodic condition caused by the topology of the M\"obius strip
are imposed on the wave function.
There are many things in the world that having spring-like structure.
For instance, DNA has a double helix structure living in our cells.
So, it is interesting to find how could such kind of helix structure would affect the behavior of a quantum field.
In fact, it is found that the  behavior of the force parallel to the axis of the helix
is very much like the force on a spring that obeys the Hooker's law in mechanics
when the ratio of the pitch to the circumference of the helix is smaller.
However, in this case, the force origins  from a quantum effect,
and so the helix structure is called a quantum spring,
see Ref.\refcite{FENG2012} for a short review.
The Casimir effect for both scalar and fermion fields under helix BCs stem from new types of space topologies was considered
\cite{Feng2010,Zhai2011,ZHAI2011a,ZHAI2011b}.
The relation between the two topologies is something like that between a cylindrical and a M\"obius strip.
The calculation of the Casimir effect under helix BCs in ($D+1$)-dimensional spacetime
shows that there is a $Z_2$ symmetry of the two space dimensions,
and that the Casimir force has a maximum value
which depends on the spacetime dimensions for both massless and massive cases.
Especially, it is shown that the Casimir force varies as the mass of the field changes.
Details of this kind of Casimir effect, will be reviewed in Sect. \ref{helix}.

Following the itinerary laid out,
we review in this paper the recent developments related to scalar Casimir effect inside a $p$-cavity mainly based on our own works.
We use the natural units $\hbar=c=k_\text{B}=1$ in this paper.

\section{The Equivalence of the Different Regularization Methods }
\label{equivreg}
\setcounter{equation}{0}
On the physical and mathematical basic of the scalar Casimir energy in a rectangular cavity,
the divergent Epstein zeta function can be reconstructed into
the form of the dual convergent Epstein zeta function plus a divergent integral
by repeated application of Abel-Plana formula,
showing explicitly the isolation of the divergence in the zeta function scheme of regularization.
Furthermore, the divergent integral can be then regulated by frequency cut-off method
and interpreted as background or geometric contribution depending on different BCs.
This investigation demonstrates that the zeta function regularization method
is identifiable with the Abel-Plana formula approach,
and it is possible that the choice of regularization methods in Casimir effect may be made for convenience.

The starting point is the energy of a massless scalar field in a rectangular cavity $\mathcal{E}=\frac12\sum_J\omega_J$.
In the case of Dirichlet or Neumann BCs,
\begin{equation}
	\mathcal{E}^\text{(D/N)}=\frac12\sum_{\vec{n}\in\mathbb{N}^D/\vec{n}\in(\mathbb{N}\cup\{\vec{0}\})^D}{'}\sqrt{(\frac{\pi n_1}{L_1})^2+(\frac{\pi n_2}{L_2})^2+\cdots+(\frac{\pi n_D}{L_D})^2},
	\label{dirichletneumann0}
\end{equation}
and in the case of periodic BCs,
\begin{equation}
	\mathcal{E}^\text{(P)}=\frac12\sum_{\vec{n}\in\mathbb{Z}^D}{'}\sqrt{(\frac{2\pi n_1}{L_1})^2+(\frac{2\pi n_2}{L_2})^2+\cdots+(\frac{2\pi n_D}{L_D})^2},
	\label{periodic0}
\end{equation}
where the superscripts ``(D), (N), (P)'' indicate the types of BCs.
And the summation, as shown in eqs.\eqref{dirichletneumann0} and \eqref{periodic0},
is over $n_1,n_2,\cdots,n_D$ from 1, 0 and $-\infty$ to $\infty$ for Dirichlet, Neumann and periodic BCs, respectively,
and the prime symbol means the case $\vec{n}=\vec{0}$ has been excluded where the vector $\vec{n}=\{n_1,\cdots,n_D\}$.
We mostly use the periodic case as the underlying example in this section,
the other two types of BCs will be briefly discussed.

\subsection{Equivalence to Abel-Plana Method in One-Dimensional Case}
In one-dimensional case,
the reflection formula of the Riemann zeta function
\begin{equation}
	\pi^{-\frac s2}\Gamma(\frac s2)\zeta(s)=\pi^{\frac{s-1}2}\Gamma(\frac{1-s}2)\zeta(1-s),
	\label{RiemannZetaReflect}
\end{equation}
which is also known as a collateral form of analytic continuation of the zeta function,
plays a key role in the regularization.
As eq.\eqref{periodic0} reduces to
\begin{equation}
	\mathcal{E}^\text{(P)}_1=\frac12\sum_{n=-\infty}^\infty\sqrt{\frac{4\pi^2n^2}{a^2}},
	\label{ep01}
\end{equation}
where $a$ is the size of the one-dimensional box,
it is quite straightforward to use eq.\eqref{RiemannZetaReflect} to obtain the regularized finite Casimir energy
\begin{equation}
	\mathcal{E}^{\text{(P),reg.}}_1=-\frac1{\pi a}\zeta(2)=-\frac{\pi}{6a}.
	\label{ep1}
\end{equation}
Although in the spirit of analytic continuation the ill-defined quantity is made equal to a finite one,
the divergency has been implicitly removed.

To show this divergency and its removal,
the regularization method using Abel-Plana formula
\begin{equation}
	\sum_{n=1}^\infty u(n)=-\frac12u(0)+\int_0^\infty u(x)\diff x+\im\int_0^\infty\frac{u(\im t)-u(-\im t)}{\e^{2\pi t}-1}\diff t
	\label{AbelPlana}
\end{equation}
is reviewed for comparison.
With eq.\eqref{AbelPlana} applied to eq.\eqref{ep01}, the first term vanishes.
The second term is $\frac{2\pi}a\int_0^\infty x^{-s}\diff x$,
which is obviously divergent for $s<0$.
One introduces the frequency cut-off function $\exp(-\delta\frac {2\pi x}a)$,
where the parameter $\delta>0$ has to be put $\delta=0$ in the end,
to illustrate the regularization and subtraction of this term.
For $s=-1$ it becomes
\begin{equation}
	\frac{2\pi}a\int_0^\infty x\e^{-\delta\frac {2\pi x}a}\diff x=\frac{a}{2\pi\delta^2}.
\end{equation}
It is proportional to the ``volume'' $a$ of the one-dimensional box,
and corresponds to the vacuum energy of the free unbounded space within the volume of the box.
The physical Casimir energy should be the difference with respect to this kind of energy,
and thus this term should be subtracted.
So what is left is the third term,
which is actually the integral form of a well-defined zeta function\cite{Schuster1999,Kurokawa2002,Schuster2005}.
In fact, for $s<0$, one can carry out the integral
\begin{equation}
\begin{split}
	\im\int_0^\infty\frac{(\im t)^{-s}-(-\im t)^{-s}}{\e^{2\pi t}-1}\diff t=&2\sin\frac{s\pi}2\int_0^\infty\frac{t^{-s}}{\e^{2\pi t}-1}\diff t\\
	=&\pi^{s-\frac12}\frac{\Gamma(\frac{1-s}2)}{\Gamma(\frac s2)}\zeta(1-s),
	\label{conjugal1}
\end{split}
\end{equation}
where the integral form of Gamma function has been used.
That is, the reflection formula of Riemann zeta function \eqref{RiemannZetaReflect} is valid only after the regularization by Abel-Plana formula.
Utilization of eq.\eqref{RiemannZetaReflect} or similar analytic continuation of Riemann zeta function
is actually implicit removal the vacuum energy of the free unbounded space within the volume of the one-dimensional box
as the Abel-Plana formula method does explicitly.

\subsection{Generalization to Higher Dimensional Cases}
For a higher dimensional case,
one uses the Epstein zeta function
\begin{equation}
	Z_D(s)=\sum_{\vec{n}\in\mathbb{Z}^D\setminus\{\vec{0}\}}(\vec{n}^2)^{-\frac s2}
	\label{EpsteinZeta}
\end{equation}
instead of the Riemann one.
The reflection formula
\begin{equation}
	\pi^{-\frac s2}\Gamma(\frac s2)Z_D(s)=\pi^{\frac{s-D}2}\Gamma(\frac{D-s}2)Z_D(D-s)
	\label{EpsteinZetaReflect}
\end{equation}
is also essential to the regularization of eq.\eqref{periodic0}.
If one chooses the box to be a hypercube with the side length of $a$,
the regularization procedure will be quite straightforward using eq.\eqref{EpsteinZetaReflect},
but the removal of divergency is also hidden.
Utilizing the results of one-dimensional case,
the proof of eq.\eqref{EpsteinZetaReflect} from the analytic continuation aspect,
the revelation of the removal of the divergency, and hence the equivalence
can be presented recursively.

It is beneficial to introduce the recurrence formula of the Epstein zeta function\cite{Lim2007},
which provides facilitation to the proof of eq.\eqref{EpsteinZetaReflect}
and the demonstration of the equivalence between the two regularization methods at length.
For homogeneous Epstein zeta function eq.\eqref{EpsteinZeta},
consider $Z_D(D-s)$, which is well-defined for $s<0$,
\begin{equation}
\begin{split}
	Z_D(D-s)=&Z_{D-1}(D-s)+2\sum_{\vec{n}\in\mathbb{Z}^{D-1}}\sum_{m\in\mathbb{N}}(\vec{n}^2+m^2)^{-\frac{D-s}2}\\
	=&Z_{D-1}(D-s)+\frac{2\pi^{\frac{D-1}2}}{\Gamma(\frac{D-s}2)}\zeta(1-s)\\
	&+\frac{4\pi^{\frac{D-s}2}}{\Gamma(\frac{D-s}2)}\sum_{\vec{n}\in\mathbb{Z}^{D-1}\setminus\{\vec{0}\}}\sum_{m\in\mathbb{N}}(\frac{\sqrt{\vec{n}^2}}m)^{\frac{1-s}2}K_{\frac{1-s}2}(2\pi m\sqrt{\vec{n}^2}),
	\label{CS1}
\end{split}
\end{equation}
where the Poisson summation formula
\begin{equation}
	\sum_{n_i=-\infty}^\infty\e^{-\frac{n_i^2\pi^2}{L_i^2}t}=\frac{L_i}{\sqrt{\pi t}}\sum_{n_i=-\infty}^\infty\e^{-\frac{n_i^2L_i^2}{t}}
	\label{poisson}
\end{equation}
and the integral form of the modified Bessel function of the second kind $K_\nu(z)$ have been used,
and the Riemann zeta term in eq.\eqref{CS1} comes from the $\vec{n}\in\{\vec{0}\}$ term.
Repeat the procedure on this recurrence formula\cite{Terras1973}, one arrives at
\begin{equation}
\begin{split}
	Z_D(D-s)=&\frac2{\Gamma(\frac{D-s}2)}\sum_{j=0}^{D-1}\pi^{\frac j2}\Gamma(\frac{D-s-j}2)\zeta(D-s-j)\\
	&+\frac{4\pi^{\frac{D-s}2}}{\Gamma(\frac{D-s}2)}\sum_{j=1}^{D-1}\sum_{\substack{m\in\mathbb{N}\\\vec{k}\in\mathbb{Z}^j\setminus\{\vec{0}\}}}(\frac{|\vec{k}|}m)^{\frac{D-s-j}2}K_{\frac{D-s-j}2}(2\pi m|\vec{k}|),
	\label{CS2}
\end{split}
\end{equation}
which as one recalls, is well-defined for $s<0$.
With eq.\eqref{CS2} and the result of one-dimensional case,
one can identify the ill-defined case of $Z_D(s),s<0$ with a finite quantity,
namely prove eq.\eqref{EpsteinZetaReflect} from the analytic continuation aspect.

Comparison to the regularization using Abel-Plana formula
is still helpful to explore the hidden removal of the divergency.
Applying eq.\eqref{AbelPlana} in $Z_D(s)$ \eqref{EpsteinZeta}, with $s<0$,
\begin{equation}
\begin{split}
	Z_D(s)=&\sum_{\vec{n}\in\mathbb{Z}^{D-1}\setminus\{\vec{0}\}}(\vec{n}^2)^{-\frac s2}+\sum_{\substack{\vec{n}\in\mathbb{Z}^{D-1}\\k\in\mathbb{Z}\setminus\{0\}}}(\vec{n}^2+k^2)^{-\frac s2}\\
	=&\sum_{\vec{n}\in\mathbb{Z}^{D-1}\setminus\{\vec{0}\}}(\vec{n}^2)^{-\frac s2}+2\sum_{\vec{n}\in\mathbb{Z}^{D-1}}\Big\{-\frac12(\vec{n}^2)^{-\frac s2}\\
	&+\int_0^\infty(\vec{n}^2+x^2)^{-\frac s2}\diff x+\im\int_0^\infty\frac{(\vec{n}^2+(\im t)^2)^{-\frac s2}-(\vec{n}^2+(-\im t)^2)^{-\frac s2}}{\e^{2\pi t}-1}\diff t\Big\}\\
	=&2\int_0^\infty(x^2)^{-\frac s2}\diff x+2\im\int_0^\infty\frac{((\im t)^2)^{-\frac s2}-((-\im t)^2)^{-\frac s2}}{\e^{2\pi t}-1}\diff t\\
	&+2\sum_{\vec{n}\in\mathbb{Z}^{D-1}\setminus\{\vec{0}\}}\int_0^\infty(\vec{n}^2+x^2)^{-\frac s2}\diff x\\
	&+2\im\sum_{\vec{n}\in\mathbb{Z}^{D-1}\setminus\{\vec{0}\}}\int_0^\infty\frac{(\vec{n}^2+(\im t)^2)^{-\frac s2}-(\vec{n}^2+(-\im t)^2)^{-\frac s2}}{\e^{2\pi t}-1}\diff t.
	\label{es4}
\end{split}
\end{equation}
On the RHS of the last equal sign,
the first term is obviously a divergent integral,
which will be canceled later.
The second term is calculated in eq.\eqref{conjugal1}.
The last term is finite and since $s<0<2$, can be carried out as
\begin{equation}
\begin{split}
	\im\int_0^\infty\frac{(\vec{n}^2+(\im t)^2)^{-\frac{s}{2}}-(\vec{n}^2+(-\im t)^2)^{-\frac{s}{2}}}{\e^{2\pi t}-1}\diff t=\frac{2\pi^{\frac s2}}{\Gamma(\frac s2)}\sum_{q\in\mathbb{N}}(\frac{|\vec{n}|}{q})^{\frac{1-s}{2}}K_\frac{1-s}{2}(2q\pi|\vec{n}|).
	\label{es44}
\end{split}
\end{equation}

The third term of eq.\eqref{es4} is still divergent.
Similar to eq.\eqref{es4},
with Abel-Plana formula \eqref{AbelPlana} employed on the summation over $\vec{n}$ once again,
this term can be written as
\begin{equation}
\begin{split}
	&2\sum_{\vec{n}\in\mathbb{Z}^{D-1}\setminus\{\vec{0}\}}\int_0^\infty(\vec{n}^2+x^2)\diff x\\
	=&-2\int_0^\infty(x^2)^{-\frac s2}\diff x+4\int_0^\infty\diff x\int_0^\infty\diff y(x^2+y^2)^{-\frac s2}\\
	&+4\im\int_0^\infty\diff x\int_0^\infty\diff t\frac{(x^2+(\im t)^2)^{-\frac s2}-(x^2+(-\im t)^2)^{-\frac s2}}{\e^{2\pi t}-1}\\
	&+4\sum_{\vec{n}\in\mathbb{Z}^{D-2}\setminus\{\vec{0}\}}\int_0^\infty\diff x\int_0^\infty\diff y(\vec{n}^2+x^2+y^2)^{-\frac s2}\\
	&+4\im\sum_{\vec{n}\in\mathbb{Z}^{D-2}\setminus\{\vec{0}\}}\int_0^\infty\diff x\int_0^\infty\diff t\frac{(\vec{n}^2+x^2+(\im t)^2)^{-\frac s2}-(\vec{n}^2+x^2+(-\im t)^2)^{-\frac s2}}{\e^{2\pi t}-1}.
	\label{es5}
\end{split}
\end{equation}
The two finite conjugal integrals of eq.\eqref{es5} can also be carried out as
\begin{equation}
\begin{split}
	4\im\int_0^\infty\diff x\int_0^\infty\diff t\frac{(x^2+(\im t)^2)^{-\frac s2}-(x^2+(-\im t)^2)^{-\frac s2}}{\e^{2\pi t}-1}=\frac{2\pi^{s-1}\Gamma(1-\frac s2)}{\Gamma(\frac s2)}\zeta(2-s),
	\label{es51}
\end{split}
\end{equation}
and
\begin{equation}
\begin{split}
	&4\im\sum_{\vec{n}\in\mathbb{Z}^{D-2}\setminus\{\vec{0}\}}\int_0^\infty\diff x\int_0^\infty\diff t\frac{(\vec{n}^2+x^2+(\im t)^2)^{-\frac s2}-(\vec{n}^2+x^2+(-\im t)^2)^{-\frac s2}}{\e^{2\pi t}-1}\\
	=&\frac{4\pi^{\frac s2}}{\Gamma(\frac s2)}\sum_{\substack{\vec{n}\in\mathbb{Z}^{D-2}\setminus\{\vec{0}\}\\q\in\mathbb{N}}}(\frac{q}{|\vec{n}|})^{\frac s2-1}K_{1-\frac s2}(s\pi q|\vec{n}|)
	\label{es52}.
\end{split}
\end{equation}
Collecting all the pieces, one has
\begin{equation}
\begin{split}
	Z_D(s)=&4\int_0^\infty\diff x\int_0^\infty\diff y(x^2+y^2)^{-\frac s2}+2\pi^{s-\frac12}\frac{\Gamma(\frac{1-s}2)}{\Gamma(\frac s2)}\zeta(1-s)\\
	&+4\sum_{\vec{n}\in\mathbb{Z}^{D-2}\setminus\{\vec{0}\}}\int_0^\infty\diff x\int_0^\infty\diff y(\vec{n}^2+x^2+y^2)^{-\frac s2}\\
	&+\frac{2\pi^{s-1}\Gamma(1-\frac s2)}{\Gamma(\frac s2)}\zeta(2-s)
	+\frac{4\pi^{\frac s2}}{\Gamma(\frac s2)}\sum_{\vec{n}\in\mathbb{Z}^{D-2}\setminus\{\vec{0}\}}\sum_{q\in\mathbb{N}}(\frac{q}{|\vec{n}|})^{\frac s2-1}K_{1-\frac s2}(s\pi q|\vec{n}|)\\
	&+\frac{2\pi^{\frac s2}}{\Gamma(\frac s2)}\sum_{\substack{\vec{n}\in\mathbb{Z}^{D-1}\setminus\{\vec{0}\}\\q\in\mathbb{N}}}(\frac{|\vec{n}|}{q})^{\frac{1-s}{2}}K_\frac{1-s}{2}(2q\pi|\vec{n}|).
	\label{es6}
\end{split}
\end{equation}

From eq.\eqref{es4} to eq.\eqref{es6},
we have seen the results of application of Abel-Plana formula \eqref{AbelPlana} once and twice.
Bit by bit, the divergency is put into the one-dimensional infinite integral in eq.\eqref{es4}
and then into the two-dimensional one in eq.\eqref{es6}
(the cancelation of the one-dimensional integral basically results from the BCs,
different situation will be discussed later),
and more and more finite conjugal integrals are isolated.
Employing Abel-Plana formula on the divergent summation and repeating the procedure for another $D-2$ times,
one then has
\begin{equation}
\begin{split}
	Z_D(s)=&2^D\int_0^\infty(x_1^2+x_2^2+\cdots+x_D^2)^{-\frac s2}\diff x_1\diff x_2\cdots\diff x_D\\
	&+\frac{\pi^{s-\frac D2}\Gamma(\frac{D-s}2)}{\Gamma(\frac s2)}\Big\{\frac2{\Gamma(\frac{D-s}2)}\sum_{j=0}^{D-1}\pi^{\frac j2}\Gamma(\frac{D-s-j}2)\zeta(D-s-j)\\
	&+\frac{4\pi^{\frac{D-s}2}}{\Gamma(\frac{D-s}2)}\sum_{j=1}^{D-1}\sum_{\substack{q\in\mathbb{N}\\\vec{n}\in\mathbb{Z}^j\setminus\{\vec{0}\}}}(\frac{|\vec{n}|}q)^{\frac{D-s-j}2}K_{\frac{D-s-j}2}(2\pi q|\vec{n}|)\Big\}.
	\label{es7}
\end{split}
\end{equation}
The terms in the brace are recognized as $Z_D(D-s)$ from the eq.\eqref{CS2}.
So finally with the help of the Abel-Plana formula,
the divergent and convergent terms of Epstein zeta function are separated as
\begin{equation}
\begin{split}
	Z_D(s)=&2^D\int_0^\infty(x_1^2+x_2^2+\cdots+x_D^2)^{-\frac s2}\diff x_1\diff x_2\cdots\diff x_D\\
	&+\frac{\pi^{s-\frac D2}\Gamma(\frac{D-s}2)}{\Gamma(\frac s2)}Z_D(D-s).
	\label{es8}
\end{split}
\end{equation}

To see more clearly what this divergent part represents,
the case that the side lengths $\{L_i,\:i=1,\cdots,D\}$ are not necessarily equal is considered.
Taking the side lengths back in eq.\eqref{es8}, for $s=-1$, the divergent part of the energy is then
\begin{equation}
	\mathcal{E}^\text{(P),div.}=2^D\pi\int_0^\infty\sqrt{(\frac{x_1}{L_1})^2+\cdots+(\frac{x_D}{L_D})^2}\diff x_1\diff x_2\cdots\diff x_D.
	\label{d2}
\end{equation}
With the frequency cut-off function similar to the one-dimensional case introduced,
this term is then regulated as
\begin{equation}
\begin{split}
	\mathcal{E}^\text{(P),div.}(\delta)=&2^D\pi\int_0^\infty\sqrt{(\frac{x_1}{L_1})^2+\cdots+(\frac{x_D}{L_D})^2}\e^{-\delta\sqrt{(\frac{2\pi x_1}{L_1})^2+\cdots+(\frac{2\pi x_D}{L_D})^2}}\diff^Dx\\
	=&\frac{\Gamma(1+D)(L_1L_2\cdots L_D)}{2^D\pi^{\frac D2}\delta^{1+D}\Gamma(\frac D2)},
	\label{d3}
\end{split}
\end{equation}
which is proportional to the volume of the $D$-dimensional box.
Just like the one-dimensional case,
this divergent term can be interpreted as the vacuum energy of the free unbounded space within the volume of the box.

Different BCs will give rise to divergent terms proportional to other geometric parameters.
In fact, following the procedure in Ref. \refcite{Lin2014},
the divergent part of the energy in the cases of Dirichlet and Neumann BCs can be expressed as
\begin{equation}
	\mathcal{E}^\text{(D/N),div.}_{(i)}=(\mp\frac12)^{D-i}\frac{\pi}2\int_0^\infty\sqrt{(\frac{x_{\mu_1}}{L_{\mu_1}})^2+\cdots+(\frac{x_{\mu_i}}{L_{\mu_i}})^2}\diff^ix,
	\label{d4}
\end{equation}
where $i=1,\cdots,D$ and $\{\mu_i\}$ is a subset of $\{1,2,\cdots,D\}$,
and the signs ``$\mp$'' correspond to Neumann and Dirichlet BCs, respectively.
Similarly, these terms are regulated with the frequency cut-off and yield
\begin{equation}
\begin{split}
	\mathcal{E}^\text{(D/N),div.}_{(i)}(\delta)=&(\mp\frac12)^{D-i}\frac{\pi}2\int_0^\infty\sqrt{(\frac{x_{\mu_1}}{L_{\mu_1}})^2+\cdots+(\frac{x_{\mu_i}}{L_{\mu_i}})^2}\e^{-\delta\sqrt{(\frac{\pi x_1}{L_1})^2+\cdots+(\frac{\pi x_D}{L_D})^2}}\diff^ix\\
	=&(\mp\frac12)^{D-i}\frac{\Gamma(i+1)(L_{\mu_1}\cdots L_{\mu_i})}{2^{i}\pi^{\frac i2}\Gamma(\frac i2)\delta^{i+1}}.
	\label{d5}
\end{split}
\end{equation}
The $i=D$ term is the same term obtained in the case of periodic BCs.
and is considered as the vacuum energy of the free unbounded space within the volume of the box.
The rest divergent terms,
which are obviously proportional to the other geometric parameters of the box,
are interpreted as the boundary or surface energy of the configuration.
In $D=2,3$ cases, this is the result obtained in Ref. \refcite{Bordag2009}.

The physical Casimir energy should be considered as the vacuum energy with these divergent terms subtracted.
When this is done, what is left in eq.\eqref{es8} can be rearranged as
\[\pi^{-\frac s2}\Gamma(\frac s2)Z_D(s)=\pi^{\frac{s-D}2}\Gamma(\frac{D-s}2)Z_D(D-s),\]
which is exactly the reflection relation of Epstein zeta function eq.\eqref{EpsteinZetaReflect}.
So the implicit riddance of divergency of zeta function technique
is prescribed by the Abel-Plana formula method of regularization.
And the two methods should be considered proven identifiable.

Through the demonstration of the equivalence of the two methods,
the structure of the divergency hidden in the analytic continuation of zeta function is shown explicitly,
which is also suggested by the the heat kernel expansion\cite{Kirsten2001,Tierz2001,Bordag2001,Nesterenko2003,Fulling2003,Bordag2009},
the well appreciated and effective analysis of the divergency of zeta function.

In the light of this equivalence,
together with their connection with other methods such as frequency cut-off\cite{Boyer1968,Svaiter1991,Svaiter1992,Beneventano1996,Butzer2011},
the consistency of using ``different'' methods to regularize different parts of the Casimir energy\cite{Geyer2008,Lin2014}
as we will do in Sect. \ref{temperature},
or to obtain different forms of the result\cite{Edery2006,Bellucci2009,Bellucci2009a},
should not be worried about.
So in the regularization of Casimir energy,
any of these methods can be chosen for convenience.

\section{Repulsive or Attractive Nature of the Casimir Force for Scalar Field}
\label{pmbox}
\setcounter{equation}{0}
The question of whether the Casimir effect for a scalar field inside a rectangular cavity
gives rise to an attractive or repulsive force has been discussed by many authors.
In this section we will re-give the general derivation
and review some investigation of this subject.
\subsection{The Casimir Energy in a $p$-Dimensional Cavity}
In Sect. \ref{equivreg}, the configuration has been set to be a hypercube for simplicity,
but in general situation, the rectangular cavities that are not closed or have unequal side lengths
may have Casimir energies and forces with different signs,
as considered in the literature\cite{Caruso1991,Li1997,Li2001}.
\subsubsection{Dirichlet and Neumann BCs}
Consider the case that in eq.\eqref{dirichletneumann0} only $p$ directions have finite side lengths,
namely in the rest $D-p$ directions, side lengths $L_i,i=p+1,p+2,\cdots,D$
can be taken to $\infty$ and the summations over these $n_i$ become integrals as
\[L_i\rightarrow\infty,\quad\frac{n_i\pi}{L_i}\rightarrow r_i,\quad\frac{\pi}{L_i}\rightarrow\diff r_i,\quad i=p+1,\cdots,D.\]
So the energy takes the form
\begin{eqnarray}
\begin{split}
	\mathcal{E}^\text{(D/N)}=&\left(\prod\limits_{i=p+1}^DL_i\right)\frac{1}{2\pi^{D-p}}\\
	&\times\sum_{\vec{n}\in\mathbb{N}^p/\vec{n}\in(\mathbb{N}\cup\{\vec{0}\})^p}\int_0^\infty\left[(\frac{\pi n_1}{L_1})^2+\cdots+(\frac{\pi n_p}{L_p})^2+r_1^2+\cdots+r_{D-p}^2\right]^{\frac12}\diff^{D-p}\mathbf{r}\\
	=&\left(\prod\limits_{i=p+1}^DL_i\right)\frac{1}{(2\sqrt{\pi})^{D-p}\Gamma(\frac{D-p}2)}\\
	&\times\sum_{\vec{n}\in\mathbb{N}^p/\vec{n}\in(\mathbb{N}\cup\{\vec{0}\})^p}\int_0^\infty r^{D-p-1}\left[(\frac{\pi n_1}{L_1})^2+\cdots+(\frac{\pi n_p}{L_p})^2+r^2\right]^{\frac12}\diff r\\.
	\label{DN1}
\end{split}
\end{eqnarray}
From Mellin transformation, the energy density is
\begin{eqnarray}
\begin{split}
	\varepsilon^\text{(D/N)}\equiv&\frac{\mathcal{E}^\text{(D/N)}}{\left(\prod\limits_{i=p+1}^DL_i\right)}\\
	=&-\frac{1}{2(2\sqrt{\pi})^{D-p+1}}\int_0^\infty t^{-\frac{D-p+3}2}\\
	&\times\left(\sum_{n_1=1/n_1=0}^\infty\e^{-\frac{n_1^2\pi^2}{L_1^2}t}\right)\cdots\left(\sum_{n_p=1/n_p=0}^\infty\e^{-\frac{n_p^2\pi^2}{L_p^2}t}\right)\diff t\\
	=&-\frac{1}{2^{D+2}\pi^{\frac{D-p+1}2}}\int_0^\infty t^{-\frac{D-p+3}2}\\
	&\times\left(\sum_{n_1=-\infty}^\infty\e^{-\frac{n_1^2\pi^2}{L_1^2}t}\mp1\right)\cdots\left(\sum_{n_p=-\infty}^\infty\e^{-\frac{n_p^2\pi^2}{L_p^2}t}\mp1\right)\diff t.
	\label{DN2}
\end{split}
\end{eqnarray}
Now the $p$ different $(\sum\e\mp1)$ factors are expanded as
\begin{eqnarray}
\begin{split}
	&\left(\sum_{n_1=-\infty}^\infty\e^{-\frac{n_1^2\pi^2}{L_1^2}t}\mp1\right)\cdots\left(\sum_{n_p=-\infty}^\infty\e^{-\frac{n_p^2\pi^2}{L_p^2}t}\mp1\right)\\
	=&\sum_{q=0}^{p-1}(\mp1)^q\sum_{\{i_1,\cdots,i_{p-q}\}\in\{1,2,\cdots,p\}}\left( \sum_{n_{i_1}=-\infty}^\infty\e^{-\frac{n_{i_1}^2\pi^2}{L_{i_1}^2}t} \right)\cdots\left( \sum_{n_{i_{p-q}}=-\infty}^\infty\e^{-\frac{n_{i_{p-q}}^2\pi^2}{L_{i_{p-q}}^2}t} \right),
	\label{binary}
\end{split}
\end{eqnarray}
where the summation $\sum_{\{i_1,\cdots,i_{p-q}\}\in\{1,2,\cdots,p\}}$
means over all the $(p-q)$-element subsets $\{i_1,\cdots,i_{p-q}\}$ of the set $\{1,2,\cdots,p\}$.
Note that if all $L_i$ are equal, eq.\eqref{binary} is just the binary expansion
\[\left(\sum_{n=-\infty}^\infty\e^{-\frac{n^2\pi^2}{L^2}t}\mp1\right)^p=\sum_{q=0}^{p-1}(\mp1)^qC_p^q\left( \sum_{n=-\infty}^\infty\e^{-\frac{n^2\pi^2}{L^2}t} \right)^{p-q}.\]
Poisson summation eq.\eqref{poisson} is the essential step of the regularization, which can be applied to all the summations.
Taking eqs.\eqref{poisson} and \eqref{binary} back into eq.\eqref{DN2} one then has
\begin{eqnarray}
\begin{split}
	\varepsilon^\text{(D/N),reg.}=&-\frac{1}{2^{D+2}\pi^{\frac{D-p+1}2}}\int_0^\infty t^{-\frac{D-p+3}2}\\
	&\times\left(\sum_{n_1=-\infty}^\infty\e^{-\frac{n_1^2\pi^2}{L_1^2}t}\mp1\right)\cdots\left(\sum_{n_p=-\infty}^\infty\e^{-\frac{n_p^2\pi^2}{L_p^2}t}\mp1\right)\diff t\\
	=&-\frac{1}{2^{D+2}\pi^{\frac{D-p+1}2}}\sum_{q=0}^{p-1}(\mp1)^q\sum_{\{i_1,\cdots,i_{p-q}\}\in\{1,2,\cdots,p\}}\frac{L_{i_1}\cdots L_{i_{p-q}}}{\pi^{\frac{p-q}2}}\\
	&\times\sum_{\vec{n}\in\mathbb{Z}^{p-q}}{'}\int_0^\infty t^{-\frac{D-q+3}2}\exp\left[ -\frac{n_{i_1}^2L_{i_1}^2+\cdots n_{i_{p-q}}^2L_{i_{p-q}}^2}{t} \right]\diff t\\
	=&-\frac{1}{2^{D+2}}\sum_{q=0}^{p-1}\frac{(\mp1)^q\Gamma(\frac{D-q+1}2)}{\pi^{\frac{D-q+1}2}}\\
	&\times\sum_{\{i_1,\cdots,i_{p-q}\}\in\{1,2,\cdots,p\}}L_{i_1}\cdots L_{i_{p-q}}Z_{p-q}(L_{i_1},\cdots,L_{i_{p-q}};D-q+1),
	\label{DN4}
\end{split}
\end{eqnarray}
where the Epstein zeta function generalized from eq.\eqref{EpsteinZeta} is defined as
\[Z_k(a_1,a_2,\cdots,a_k;s)\equiv\sum_{\vec{n}\in\mathbb{Z}^k\setminus\{\vec{0}\}}(a_1^2n_1^2+\cdots+a_k^2n_k^2)^{-\frac s2}.\]
Eq.\eqref{DN4} is the general form of the regularized Casimir energy (density) in a rectangular cavity with Dirichlet or Neumann BCs.
The ``$-$'' sign in $(\mp1)^q$ corresponds to Dirichlet BCs and ``$+$'' sign to Neumann BCs.

In two-dimensional closed box with Dirichlet BCs, i.e. $D=p=2$, let $L_1=a$, $L_2=b$, eq.\eqref{DN4} is
\begin{eqnarray}
\begin{split}
	\mathcal{E}_2^\text{(D),reg.}=&-\frac{ab}{32\pi}Z_2(a,b;3)+\frac{\pi}{48}(\frac1a+\frac1b).
	\label{D2d}
\end{split}
\end{eqnarray}
And in three-dimensional closed box with Dirichlet BCs, i.e. $D=p=3$, let $L_1=a$, $L_2=b$, $L_3=c$, eq.\eqref{DN4} is
\begin{eqnarray}
\begin{split}
	\mathcal{E}_3^\text{(D),reg.}=&-\frac{abc}{32\pi^2}Z_3(a,b,c;4)+\frac{bc}{64\pi}Z_2(b,c;3)+\frac{ac}{64\pi}Z_2(a,c;3)\\
	&+\frac{ab}{64\pi}Z_2(a,b;3)-\frac{\pi}{96}(\frac1a+\frac1b+\frac1c).
	\label{D3d}
\end{split}
\end{eqnarray}
Both eqs.\eqref{D2d} and \eqref{D3d} have been obtained in Refs. \refcite{Actor1994,Actor1995} and reviewed in Ref. \refcite{Bordag2009}.

In the case that the cavity has equal edges $L_i=L,i=1,\cdots,p$ with Dirichlet BCs, eq.\eqref{DN4} becomes
\begin{equation}
	\varepsilon^\text{(D),reg.}=-\frac{L^{p-D-1}}{2^{D+2}}\sum_{q=0}^{p-1}C_p^q\frac{(\mp1)^q\Gamma(\frac{D-q+1}2)}{\pi^{\frac{D-q+1}2}}Z_{p-q}(1,\cdots,1;D-q+1),
	\label{DN5}
\end{equation}
which is what Caruso et al. have obtained in Ref. \refcite{Caruso1991}.

Taking ``$+$'' in every ``$\mp$'' sign, the energy (density) in Neumann case is always negative.
However, since there is a factor of $(-1)^q$ in eq.\eqref{DN4} for Dirichlet BCs,
the sign of the energy (density) is not yet determinative.
In Ref. \refcite{Caruso1991} the authors analysed the case in which all the edges are equal.
If $p$ is odd, it can be seen analytically that the energy (density) is also negative.
If $p$ is even, it is found out numerically that for every even $p$ there is a critical $D=D_\text{c}$,
and the energy (density) is positive for $D<D_\text{c}$.

As for the more general case with unequal sidelengths,
Ref. \refcite{Li1997} provides an angle to address this issue.
Accroding to definition, the energy density in $p$-cavity $\varepsilon_p^\text{(D)}$ with Dirichlet BCs has
\begin{equation}
	\mathcal{E}^\text{(D)}=\left( \prod_{i=p+1}^DL_i \right)\varepsilon_p^\text{(D)},\:\text{for }L_{p+1},\cdots,L_D\gg L_1,\cdots,L_p.
	\label{down1}
\end{equation}
It follows that
\begin{equation}
	\varepsilon_p^\text{(D)}=L_{p-q+1}L_{p-q+2}\cdots L_p\varepsilon_{p-q}^\text{(D)},\:\text{for }L_{p-q+1},\cdots,L_p\gg L_1,\cdots,L_{p-q}.
	\label{down2}
\end{equation}
Eqs.\eqref{down1} and \eqref{down2} are also valid for regularized energy densities.
This shows that the ratio of side lengths may have a critical value for energy density to change sign.
For example, $\varepsilon_2^\text{(D),reg.}=L_2\varepsilon_1^\text{(D),reg.}$ for $L_2\gg L_1$.
Now $\varepsilon_1^\text{(D),reg.}$ is always negative
and $\varepsilon_2^\text{(D),reg.}(L_1=L_2)$ is positive for $D<D_\text{crit}=6$\cite{Caruso1991}.
Since $\varepsilon_2^\text{(D),reg.}(L_1,L_2)$ is a continuous function for $L_2>0$,
there thus exists a critical ratio $\mu=L_2/L_1=\mu_\text{crit}$ for $\varepsilon_2^\text{(D),reg.}$ to turn from positive to negative.
Furthermore, there is a $Z_2$ symmetry $L_1\leftrightarrow L_2$ for energy function.
Numerical calculations\cite{Li1997} show that for $p=2$ and $D<D_\text{crit}=6$\cite{Caruso1991},
the critical ratio $\mu_\text{crit}$ does exists.
When $D=2=p$, $\mu_\text{crit}=2.737$, and as $D$ increases $\mu_\text{crit}$ becomes smaller.
And one can conclude that if $L_1/L_2>2.737$ or $L_2/L_1>2.737$,
the energy density $\varepsilon_p^\text{(D,reg.)}<0$ for any space dimensionality $D$.
\subsubsection{Periodic BCs}
In periodic case eq.\eqref{periodic0}, $L_i,i=p+1,\cdots,D$ are also taken to $\infty$
and summations over these $n_i$ are turned into integrals with
\[L_i\rightarrow\infty,\quad\frac{2n_i\pi}{L_i}\rightarrow r_i,\quad\frac{2\pi}{L_i}\rightarrow\diff r_i,\quad i=p+1,\cdots,D.\]
And then similarly, energy density is
\begin{eqnarray}
\begin{split}
	\varepsilon^\text{(P)}\equiv&\frac{\mathcal{E}^\text{(P)}}{\left(\prod\limits_{i=p+1}^DL_i\right)}\\
	=&\frac{1}{2^{D-p}\pi^{\frac{D-p}2}\Gamma(\frac{D-p}2)}\sum_{\vec{n}\in\mathbb{Z}^p}\int_0^\infty r^{D-p-1}\left[(\frac{2\pi n_1}{L_1})^2+\cdots+(\frac{2\pi n_p}{L_p})^2+r^2\right]^{\frac12}\diff r\\
	=&-\frac{1}{2(2\sqrt{\pi})^{D-p+1}}\int_0^\infty t^{-\frac{D-p+3}2}\left( \sum_{n_1=-\infty}^\infty\e^{-4\frac{n_1^2\pi^2}{L_1^2}t} \right)\cdots\left( \sum_{n_p=-\infty}^\infty\e^{-4\frac{n_p^2\pi^2}{L_p^2}t} \right)\diff t.
	\label{P2}
\end{split}
\end{eqnarray}
Again, the Poisson summation eq.\eqref{poisson} is employed for regularization,
\begin{eqnarray}
\begin{split}
	\varepsilon^\text{(P),reg.}=&-\frac{1}{2(2\sqrt{\pi})^{D-p+1}}\int_0^\infty t^{-\frac{D-p+3}2}\\
	&\times\left( \frac{L_1}{2\sqrt{\pi t}}\sum_{n_1=-\infty}^\infty\e^{-\frac{n_1^2L_1^2}{4t}} \right)\cdots\left( \frac{L_p}{2\sqrt{\pi t}}\sum_{n_p=-\infty}^\infty\e^{-\frac{n_p^2L_p^2}{4t}} \right)\diff t\\
	=&-\frac{L_1\cdots L_p}{2(2\sqrt{\pi})^{D+1}}\sum_{\vec{n}\in\mathbb{Z}^p}{'}\int_0^\infty t^{-\frac{D+3}2}\e^{-(n_1^2L_1^2+\cdots n_p^2L_p^2)\frac1{4t}}\diff t\\
	=&-\frac{L_1\cdots L_p}{2\pi^{\frac{D+1}2}}\Gamma(\frac{D+1}2)Z_p(L_1,\cdots,L_p;D+1).
	\label{P4}
\end{split}
\end{eqnarray}
Eq.\eqref{P4} is the general form of the regularized Casimir energy (density) in a rectangular cavity with periodic BCs,
which is in agreement of what Ambj{\o}rn and Wolfram have obtained\cite{Ambjorn1983},
and obvious is always negative.
In two and three-dimensional closed box with periodic BCs, i.e. $D=p=2,3$,
eq.\eqref{P4} recovers the results obtained in Refs. \refcite{Actor1994,Actor1995} and reviewed in Ref. \refcite{Bordag2009}.

As a summary, we have put all the results reviewed above in Table \ref{pmboxtable2}.
\begin{table}[!htp]
	\centering	
	\tbl{Sign of the Casimir energy density of a massless scalar field confined in $p$-cavity of $(D+1)$-dimensional spacetime}
	{\begin{tabular}{c|c|c|cc}
		\hline
		\hline
		\multirow{2}{*}{$\varepsilon^\text{reg.}$}	&\multicolumn{2}{c|}{equal $p$ edges}	&\multicolumn{2}{c}{\multirow{2}{*}{unequal $p$ edges}}\\
		\cline{2-3}
								&$p$ odd	&$p$ even		&\\
		\hline
		periodic&\multicolumn{4}{c}{$<0$}\\
		\hline
		Neumann&\multicolumn{4}{c}{$<0$}\\
		\hline
		\multirow{2}{*}{Dirichlet}&\multirow{2}{*}{$<0$}	&$>0$ for $D<D_\text{c}$	&\multicolumn{2}{c}{depends on $p$, $D$}\\
		\cline{3-3}
		&&$<0$ for $D>D_\text{c}$&\multicolumn{2}{c}{and the ratios of sidelengths}\\
		\hline
	\end{tabular}}
	\label{pmboxtable2}
\end{table}

\subsection{The Sign of the Force}
From eqs.\eqref{DN4} and \eqref{P4}, one can calculate the Casimir force per unit area for a specific $p$.
For $p=1$,
\begin{equation}
	\varepsilon_1^\text{(D/N),reg.}=-\frac{\Gamma(\frac{D+1}2)\zeta(D+1)}{2^{D+1}\pi^{\frac{D+1}2}L^D},
	\label{epsilonp=1}
\end{equation}
and $\varepsilon_1^\text{(P),reg.}=2^{D+1}\varepsilon_1^\text{(D/N),reg.}$.
So the Casimir force density for $p=1$ is always negative, namely attractive for all three types of BCs.

In the $p=2$ case, one can expand the Epstein zeta function $Z_2(L_1,L_2;D+1)$ in the similar manner as eq.\eqref{CS2}
and rewrite the energy density eq.\eqref{DN4} for Dirichlet BCs as
\begin{equation}
	\begin{split}
	\varepsilon_2^\text{(D),reg.}=&-\frac1{2^{D-1}L_1^{\frac D2}L_2^{\frac D2-1}}\sum_{n_1,n_2=1}^\infty\left( \frac{n_1}{n_2} \right)^{\frac D2}K_\frac D2\left( 2n_1n_2\pi\frac{L_2}{L_1} \right)\\
	&-\frac{\zeta(D+1)\Gamma(\frac{D+1}2)L_2}{2^{D+1}\pi^{\frac{D+1}2}L_1^D}+\frac{\zeta(D)\Gamma(\frac D2)}{2^{D+1}\pi^{\frac D2}L_1^{D-1}}.
	\end{split}
	\label{epsilonDp=2}
\end{equation}

It follows that the force density along the direction of $L_2$ is
\begin{equation}
	\begin{split}
		-\frac{\partial\varepsilon_2^\text{(D),reg.}}{\partial L_2}=&\frac{\left( 1-\frac D2 \right)}{2^{D-1}\left( L_1L_2 \right)^{\frac D2}}\sum_{n_1,n_2=1}^\infty\left( \frac{n_1}{n_2} \right)^{\frac D2}K_\frac D2\left( 2n_1n_2\pi\frac{L_2}{L_1} \right)\\
		&+\frac{\pi}{2^{D-2}L_1^{\frac D2+1}L_2^{\frac D2-1}}\sum_{n_1,n_2=1}^\infty\frac{n_1^{\frac D2+1}}{n_2^{\frac D2-1}}K'_\frac D2\left( 2n_1n_2\pi\frac{L_2}{L_1} \right)\\
		&+\frac{\zeta(D+1)\Gamma(\frac{D+1}2)}{2^{D+1}\pi^{\frac{D+1}2}L_1^D},
	\end{split}
	\label{forceDp=2}
\end{equation}
where $K'_\nu(z)\equiv\frac{\partial K_\nu(z)}{\partial z}$.
The first two terms in eq.\eqref{forceDp=2} are negative and monotonically increasing functions with $L_2$,
while the last term is positive and is independent of $L_2$.
So it is expected that the Casimir energy density has a maximum with respect to $L_2$,
and that the Casimir force density will turn from attractive to repulsive when $L_2$ increases.
\begin{table}[!htp]
	\centering	
	\tbl{The maximum value of the Casimir energy densities at $L_2/L_1=\mu_\text{max}$ for massless scalar fields
satisfying Diriclet BCs inside a cavity with unequal edges in a $(D+1)$-dimensional spacetime,
where $L_1$ is the chosen unit length.
Meantime, the values of $\varepsilon_2^\text{(D),reg.}$ at $L_1=L_2$ are listed to contrast with $\varepsilon_{2,\text{max}}^\text{(D),reg.}$.}
	{\begin{tabular}{cccc}
		\hline
		\hline
		$D$&$\mu_\text{max}$&$\varepsilon_{2,\text{max}}^\text{(D),reg.}$&$\varepsilon_2^\text{(D),reg.}(L_1=L_2)$\\
		\hline
		5&$1+(1\times10^{-7})$&0.0001146407&0.0001146408\\
		6&1.0102&-0.0000192394&-0.0000194771\\
		7&1.0375&-0.0000366757&-0.0000386962\\
		8&1.0575&-0.0000311072&-0.0000341599\\
		9&1.0724&-0.0000231299&-0.0000263762\\
		10&1.0830&-0.0000167097&-0.0000197328\\
		11&1.0911&-0.0000121189&-0.0000147795\\
		12&1.0968&-0.0000089401&-0.0000112286\\
		13&1.1008&-0.0000067468&-0.0000087042\\
		14&1.1034&-0.0000052212&-0.0000069027\\
		15&1.1049&-0.0000041471&-0.0000056059\\
		16&1.1058&-0.0000033808&-0.0000043828\\
		17&1.1063&-0.0000028276&-0.0000039731\\
		18&1.1065&-0.0000024248&-0.0000034650\\
		19&1.1067&-0.0000021304&-0.0000030916\\
		\hline
	\end{tabular}}
	\label{pmboxtable1}
\end{table}
Table \ref{pmboxtable1}\cite{Li1997} shows the maximum of the energy density for $p=2$ and various $D$ with Dirichlet BCs
and the corresponding ratios of side lengths $\mu_\text{max}$,
as well as the energy density at $L_1=L_2$ for contrast.

Similar analyses can be applied to eq.\eqref{DN4} of Neumann case and eq.\eqref{P4} of periodic case.
It is found that although the energy densities with these two types of BCs is always negative,
the force densities show similar behaviors as in Dirichlet case,
i.e. they turn from attractive to repulsive as $L_2$ increases.

For higher dimensional case, if one considers the force along only the direction of $L_p$,
namely $-\partial \varepsilon_p^\text{reg.}/\partial L_p$
and let all the rest side lengths be equal $L_1=L_2=\cdots=L_{p-1}$,
the analyses of 2-dimensional case above can be extended straightforward
and the conclusion is still valid.

These results permit us to discuss a possible application for the Abraham-Lorentz electron model\cite{Li1997}.
A $p=3$ rectangular cavity with walls of perfect conductivity is considered.
The electrmagnetic field satisfies the BCs $\mathbf{n}\cdot\mathbf{B}=0$ and $\mathbf{n}\cdot\mathbf{E}=0$.
The Casimir energy $\mathcal{E}^\text{(em)}(L_1,L_2,L_3)$ of the electromagnetic field can be
written in terms of the massless scalar field as
\begin{equation}
	\mathcal{E}^\text{(em)}(L_1,L_2,L_3)=2\mathcal{E}^\text{(D)}(L_1,L_2,L_3)+\mathcal{E}^\text{(D)}(L_1,L_2)+\mathcal{E}^\text{(D)}(L_1,L_3)+\mathcal{E}^\text{(D)}(L_2,L_3).
	\label{em}
\end{equation}
Use will also be made of the well-known fact\cite{Peterson1982} that
the order of magnitude of the electromagnetic zero-point energy does not change
if one deforms a spherical shell of radius $a$ into a cubic shell of length, with $L\approx2a$.
On the other hand, the Abraham-Lorentz model describes the electron as
a conducting spherical shell of radius $a$.
To guarantee the stability of the electron Poincar{\'e} stresses had to be postulated.
Casimir\cite{Casimir1953} proposed to extend the classical electron model
by taking into account the zero-point fluctuations of the electromagnetic field
inside and outside of the conducting shell.
Unfortunately, the Casimir model of the electron fails,
at least in the $L_1=L_2=L_3\approx2a$ case,
because the Casimir energy of an $S^2$ electron is positive from eq.\eqref{em}.
Does this argument still hold for rectangular cavity?
The answer is no, and it can be shown that the zero-point energy is negative
when we choose lengths of edges, appropriately.
We take, for example $L_1=1.6$ and $L_2=L_3=1$,
then $\mathcal{E}^\text{(em)}(L_1,L_2,L_3)\approx-2\times10^{-3}$.
Therefore, Casimir-like model of electron could be stable.
Note that, in this case, the condition of stability will be satisfied only for a particular shape and size.

\section{Casimir Pistons}
\label{piston}
\begin{figure}[htp!]
	\centering
	\includegraphics[width=0.7\textwidth]{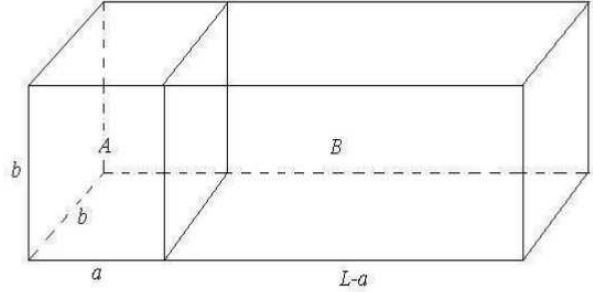}
	\caption{Three-dimensional piston with the variable edges of $a$ and $L-a$ and fixed edges of $b$.}
	\label{pistonfig1}
\end{figure}

\setcounter{equation}{0}
The rectangular Casimir piston is a variant of the Casimir effect in a rectangular box,
where the box is divided by a movable partition.
Based on the Casimir energy reviewed in Sect. \ref{pmbox}, for simplicity,
we will focus on the configuration of a $p$-dimensional (note that the dimensionality of the space $D=p$) rectangular piston
where the length of the variable edges are $a$ and $L-a$ while the rest sidelengths are all $b$.
Fig.\ref{pistonfig1} illustrates a three-dimensional piston.
It follows that the Casimir force on the piston is
\begin{equation}
	\mathcal{F}_p=-\frac{\partial}{\partial a}\left[ \mathcal{E}_p(A)+\mathcal{E}_p(B)+\mathcal{E}^\text{out} \right]=-\frac{\partial}{\partial a}\left[ \mathcal{E}_p(A)+\mathcal{E}_p(B) \right],
	\label{f}
\end{equation}
where $\mathcal{E}_p(A)$ and $\mathcal{E}_p(B)$ are the Casimir energy in compartment A and B indicated in Fig.\ref{pistonfig1}, respectively,
and $\mathcal{E}^\text{out}$ is the energy outside the box,
which provides no contribution to the force and will be neglected in the following.

\subsection{Massless Scalar Field with Dirichlet BCs}
For Dirichlet BCs, the energy of a massless scalar field in a rectangular cavity is given in eq.\eqref{DN4}.
The case that $D=p$ and with ``$-$'' sign in ``$\mp$'' is retained here:
\begin{equation}
	\begin{split}
	\mathcal{E}_p^\text{(D),reg.}=&-\frac{1}{2^{p+2}}\sum_{q=0}^{p-1}\frac{(-1)^q\Gamma(\frac{p-q+1}2)}{\pi^{\frac{p-q+1}2}}\\
	&\times\sum_{\{i_1,\cdots,i_{p-q}\}\in\{1,2,\cdots,p\}}L_{i_1}\cdots L_{i_{p-q}}Z_{p-q}(L_{i_1},\cdots,L_{i_{p-q}};p-q+1).
	\end{split}
	\label{eDcavity}
\end{equation}
For compartment A, the energy is expressed as
\begin{equation}
	\begin{split}
	\mathcal{E}_p^\text{(D),reg.}(a,b)=&-\frac a{2^{p+2}}\sum_{q=0}^{p-1}\frac{(-1)^q\Gamma(\frac{p-q+1}2)}{\pi^{\frac{p-q+1}2}}\\
	&\times C_{p-1}^qb^{p-q-1}\sum_{(j,\vec{k})\in\mathbb{Z}^{p-q}}{'}\left[ a^2j^2+b^2\vec{k}^2 \right]^{-\frac{p-q+1}2}+c(b),
	\end{split}
	\label{eDa}
\end{equation}
where similarly $\vec{k}$ represents $\{k_1,\cdots,k_{p-1}\}$
and the last term $c(b)$ contains terms independent of $a$ and thus provides no contribution to the force on the piston.

For $p=1$, the energy is simply
\begin{equation}
	\mathcal{E}^\text{(D),reg.}_1(a)=-\frac{\zeta(2)}{4\pi a}=-\frac{\pi}{24a}.
	\label{eDp1}
\end{equation}
Thus the force on the piston is
\begin{equation}
	\mathcal{F}^\text{(D)}_1(a)=-\lim_{L\rightarrow\infty}\frac{\partial}{\partial a}\left[ \mathcal{E}^\text{(D)}_1(a)+\mathcal{E}^\text{(D)}_1(L-a) \right]=-\frac{\pi}{24a^2}.
	\label{fDp1}
\end{equation}

For $p\ge2$, eq.\eqref{eDa} is then manipulated as follows
\begin{equation}
	\begin{split}
		\mathcal{E}_p^\text{(D),reg.}(a,b)=&-\frac a{2^{p+2}}\sum_{q=0}^{p-2}\frac{(-1)^q\Gamma(\frac{p-q+1}2)}{\pi^{\frac{p-q+1}2}}C_{p-1}^qb^{p-q-1}\\
		&\times\sum_{j=-\infty}^\infty{'}\sum_{\vec{k}\in\mathbb{Z}^{p-q-1}}\left[ a^2j^2+b^2\vec{k}^2 \right]^{-\frac{p-q+1}2}\\
		&-\frac a{2^{p+2}b^2}\sum_{q=0}^{p-2}(-1)^qC_{p-1}^qZ_{p-q-1}(p-q+1)+\frac{(-1)^p}{2^{p+1}\pi a}\zeta(2)+c(b)\\
		=&-\frac 1{2^pb}\sum_{q=0}^{p-2}(-1)^qC_{p-1}^q\sum_{j=1}^\infty\sum_{\vec{k}\in\mathbb{Z}^{p-q-1}}{'}\frac{|\vec{k}|}jK_1(2\frac abj|\vec{k}|\pi)\\
		&-\frac a{2^{p+2}b^2}\sum_{q=0}^{p-2}(-1)^qC_{p-1}^qZ_{p-q-1}(p-q+1)+c(b),
	\end{split}
	\label{eDa2}
\end{equation}
where the Poisson summation eq.\eqref{poisson}
and the integral form of the modified Bessel function of the second kind have come to one's aid,
and the homogeneous Epstein zeta terms are the $j=0$ terms.

Now the force on the piston is
\begin{equation}
	\begin{split}
		\mathcal{F}_p^\text{(D)}=&-\lim_{L\rightarrow\infty}\frac{\partial}{\partial a}\left[ \mathcal{E}_p^\text{(D),reg.}(a,b)+\mathcal{E}_p^\text{(D),reg.}(L-a,b) \right]\\
		=&\frac{\pi}{2^{p-1}b^2}\sum_{q=0}^{p-2}(-1)^qC_{p-1}^q\sum_{j=1}^\infty\sum_{\vec{k}\in\mathbb{Z}^{p-q-1}}{'}|\vec{k}|^2K_1'(2\frac abj|\vec{k}|\pi),
	\end{split}
	\label{fD}
\end{equation}
where $K_1'(x)=\frac{\partial}{\partial x}K_1(x)$.

When $p=2$,
\begin{equation}
	\mathcal{F}_2^\text{(D)}=\frac{\pi}{b^2}\sum_{j,k=1}^\infty k^2K_1'(2\frac abjk\pi),
	\label{fD2}
\end{equation}
which is the result obtained by Cavalcanti\cite{Cavalcanti2004}.

\subsection{Massive Scalar Field with Dirichlet BCs}
On the other hand, the Casimir effect for the massive scalar field also studied by some authors\cite{Bordag2001,Pinto2003,Barone2004}.
As is known that the Casimir effect vanishes as the mass $m$ of the field goes to infinity
since there are no more quantum fluctuations in the limit.
We review here the study of the precise way the Casimir energy varies as the mass changes.
Since the regularized energy for massive case with mass $m$ is not given in previous sections,
in order to be self-contained, we start from the unregularized energy in compartment A:
\begin{equation}
	\mathcal{E}_p^\text{(D)}(a,b,m)=\frac12\sum_{j=1}^\infty\sum_{k_1,\cdots,k_{p-1}=1}^\infty\sqrt{\frac{\pi^2j^2}{a^2}+\frac{\pi^2\vec{k}^2}{b^2}+m^2}.
	\label{eDma}
\end{equation}
The regularization procedure is similar, the Mellin transformation and Poisson summation eq.\eqref{poisson} are employed on the summations.
\begin{equation}
	\begin{split}
		\mathcal{E}_p^\text{(D)}(a,b,m)=&-\frac1{4\sqrt{\pi}}\sum_{j=1}^\infty\sum_{k_1,\cdots,k_{p-1}=1}^\infty\int_0^\infty t^{-\frac32}\e^{-\left[ \frac{j^2\pi^2}{a^2}+\frac{\vec{k}^2\pi^2}{b^2}+m^2 \right]t}\diff t\\
		=&-\frac1{2^{p+2}\sqrt{\pi}}\int_0^\infty t^{-\frac32}\left( \sum_{j=-\infty}^\infty\e^{-\frac{j^2\pi^2}{a^2}t}-1 \right)\\
		&\times\left( \sum_{k=-\infty}^\infty\e^{-\frac{k^2\pi^2}{b^2}t}-1 \right)^{p-1}\e^{-m^2t}\diff t\\
		=&-\frac1{2^{p+2}\sqrt{\pi}}\sum_{q=0}^{p-1}(-1)^qC_{p-1}^q\int_0^\infty t^{-\frac32}\e^{-m^2t}\\
		&\times\left( \sum_{j=-\infty}^\infty\e^{-\frac{j^2\pi^2}{a^2}t} \right)\left( \sum_{\vec{k}\in\mathbb{Z}^{p-q-1}}\e^{-\frac{\vec{k}^2\pi^2}{b^2}t} \right)\diff t+c(b),
	\end{split}
	\label{eDma1}
\end{equation}
where all the $a$ independent terms are still put in $c(b)$, then the regularized energy
\begin{equation}
	\begin{split}
		&\mathcal{E}_p^\text{(D),reg.}(a,b,m)\\
		=&-\frac a{2^p}\sum_{q=0}^{p-2}(-1)^qC_{p-1}^q\frac{b^{p-q-1}}{\pi^{\frac{p-q+1}2}}\sum_{j=1}^\infty\sum_{\vec{k}\in\mathbb{Z}^{p-q-1}}\left( \frac{m^2}{j^2a^2+\vec{k}^2b^2} \right)^{\frac{p-q+1}4}\\
		&\times K_{\frac{p-q+1}2}\left(2m\sqrt{j^2a^2+\vec{k}^2b^2}  \right)\\
		&-\frac a{2^{p+1}}\sum_{q=0}^{p-2}(-1)^qC_{p-1}^q\frac{b^{p-q-1}}{\pi^{\frac{p-q+1}2}}\sum_{\vec{k}\in\mathbb{Z}^{p-q-1}}\left( \frac{m^2}{\vec{k}^2b^2} \right)^{\frac{p-q+1}4}K_{\frac{p-q+1}2}\left( 2m|\vec{k}|b \right)\\
		&+\frac{(-1)^p}{2^p\pi}\sum_{j=1}^\infty\frac mjK_1\left( 2mja \right)+\frac{(-1)^pa\Gamma(-1)m^2}{2^{p+2}\pi}+c(b).
	\end{split}
	\label{eDma3}
\end{equation}
And the force yields
\begin{equation}
	\begin{split}
		\mathcal{F}_p^\text{(D)}(a,b,m)=&-\frac1{2^p}\sum_{q=0}^{p-2}(-1)^qC_{p-1}^q\frac{b^{p-q-1}}{\pi^{\frac{p-q+1}2}}\\
		&\times\sum_{j=1}^\infty\sum_{\vec{k}\in\mathbb{Z}^{p-q-1}}\left[ \frac{m^{\frac{p-q+1}2}}{\left( a^2j^2+b^2\vec{k}^2 \right)^{\frac{p-q+1}4}}K_{\frac{p-q+1}2}\left( 2\sqrt{a^2j^2+b^2\vec{k}^2}m \right)\right.\\
		&\left.-\frac{m^{\frac{p-q+3}2}}{\left( a^2j^2+b^2\vec{k}^2 \right)^{\frac{p-q+3}4}}K_{\frac{p-q+3}2}\left( 2\sqrt{a^2j^2+b^2\vec{k}^2}m \right) \right]\\
		&-\frac{(-1)^p}{2^p\pi a}\sum_{j=1}^\infty\frac mjK_1\left( 2mja \right)+\frac{(-1)^pm^2}{2^{p-1}\pi a}\sum_{j=1}^\infty K_2\left( 2mja \right).
	\end{split}
	\label{fDm}
\end{equation}

\begin{figure}[htp!]
	\centering
	\includegraphics[width=0.5\textwidth]{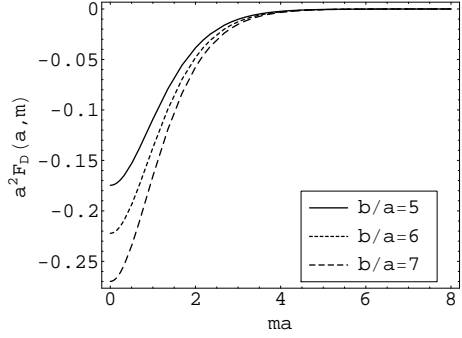}
	\caption{The Casimir force on the piston (in units of $\frac1{a^2}$) versus $ma$ for different ratio of $b/a$ with Dirichlet BCs.}
	\label{pistonfig2}
\end{figure}

\begin{figure}[htp!]
	\centering
	\includegraphics[width=0.5\textwidth]{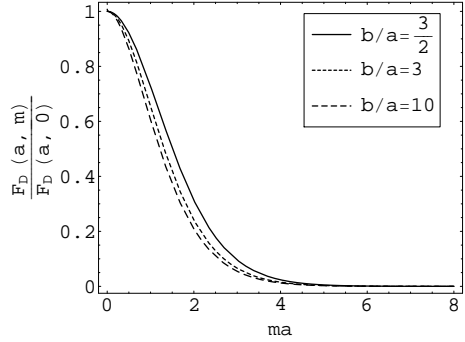}
	\caption{The ratio of the Casimir force on the piston for a massive scalar field and massless scalar field versus $ma$
		for different ratio of $b/a$ with Dirichlet BCs.}
	\label{pistonfig3}
\end{figure}

For $p=2$, eqs.\eqref{eDma3} and \eqref{fDm} give the results obtained in Ref. \refcite{ZHAI2009},
in which the authors also consider the influence of the mass on the force illustrated in Figs.\ref{pistonfig2} and \ref{pistonfig3}.

\subsection{Hybrid BCs}
The same configuration as Fig. \ref{pistonfig1} indicates is considered,
except that the BC on the piston is Neumann while those on the rest are Dirichlet,
which is dubbed as the hybrid BCs and denoted in the superscript ``(H)''.

The energy in compartment A with hybrid BCs, before regularization, is
\begin{equation}
	\mathcal{E}^\text{(H)}_p(a,b)=\frac12\sum_{j=0}^\infty\sum_{k_1,\cdots,k_{p-1}=1}^\infty\sqrt{(j+\frac12)^2\frac{\pi^2}{a^2}+\frac{\vec{k}^2b^2}{b^2}},
	\label{eH}
\end{equation}
for massless case and
\begin{equation}
	\mathcal{E}^\text{(H)}_p(a,b,m)=\frac12\sum_{j=0}^\infty\sum_{k_1,\cdots,k_{p-1}=1}^\infty\sqrt{(j+\frac12)^2\frac{\pi^2}{a^2}+\frac{\vec{k}^2b^2}{b^2}+m^2},
	\label{eHm}
\end{equation}
for massive case.
One can re-express eqs.\eqref{eH} and \eqref{eHm} as\cite{Zhai2007}
\begin{eqnarray}
	\label{eH1}
	\mathcal{E}^\text{(H)}_p(a,b)=&\mathcal{E}^\text{(D)}_p(2a,b)-\mathcal{E}^\text{(D)}_p(a,b),\\
	\label{eHm1}
	\mathcal{E}^\text{(H)}_p(a,b,m)=&\mathcal{E}^\text{(D)}_p(2a,b,m)-\mathcal{E}^\text{(D)}_p(a,b,m).
\end{eqnarray}
Therefore, the force on the piston with hybrid BCs can be obtained by
\begin{eqnarray}
	\label{fH}
	\mathcal{F}^\text{(H)}_p(a,b)=&2\mathcal{F}^\text{(D)}_p(2a,b)-\mathcal{F}^\text{(D)}_p(a,b),\\
	\label{fHm}
	\mathcal{F}^\text{(H)}_p(a,b,m)=&2\mathcal{F}^\text{(D)}_p(2a,b,m)-\mathcal{F}^\text{(D)}_p(a,b,m).
\end{eqnarray}

For $p=1$ and massless case, from eq.\eqref{fDp1},
\begin{equation}
	\mathcal{F}_1^\text{(H)}(a)=2\mathcal{F}^\text{(D)}_1(2a)-\mathcal{F}^\text{(D)}_1(a)=\frac{\pi}{48a^2},
	\label{fHp1}
\end{equation}
which was obtained in Ref. \refcite{Zhai2007}, and can also be obtained by use of exponential cutoff technique\cite{Fulling2007}.

\begin{figure}[htp!]
	\centering
	\includegraphics[width=0.5\textwidth]{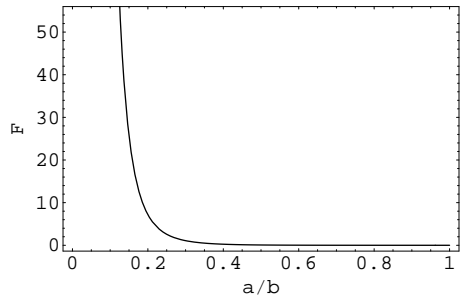}
	\caption{Casimir force $\mathcal{F}$ (in units $\hbar c/b^2$) on a three-dimensional piston
	versus $a/b$ where $a$ is the plate separation
	and $b$ is the length of the sides of the square base.}
	\label{pistonfig4}
\end{figure}

For $p\ge2$, from eqs.\eqref{fD} one can obtain
\begin{equation}
	\mathcal{F}_p^\text{(H)}(a,b)=\frac{\pi}{2^{p-1}b^2}\sum_{q=0}^{p-2}(-1)^qC_{p-1}^q\sum_{j=1}^\infty\sum_{\vec{k}\in\mathbb{Z}^{p-q-1}}{'}|\vec{k}|^2\left[ 2K_1'(4\frac abj|\vec{k}|\pi)-K_1'(2\frac abj|\vec{k}|\pi)\right],
	\label{fH1}
\end{equation}
for massless case, and for $p=2,3$, it is the result obtained in Ref. \refcite{Zhai2007},
in which numerical computation has been carried out for all $p=1,2,3$ cases showing that the force is always repulsive (see Fig.\ref{pistonfig4} for $p=3$),
in contrast with the same problem where the BCs are Dirichlet on all surfaces.
For the massive case, the influence of the mass is similar to the Dirichlet BCs (see Ref. \refcite{Zhai2007}).

The problem of hybrid BCs reviewed here is analogous
to the problem in electromagnetic field that the piston is an
infinitely permeable plate and the other sides of the cavity are
perfectly conducting ones. This problem may be connected with the study of
dynamical Casimir effect and may be applied to the fabrication of
microelectromechanical system (MEMS).

\section{Nonzero Temperature Casimir Effect}
\label{temperature}
\setcounter{equation}{0}
In previous sections, the Casimir effect of a quantum field on a vacuum state has been reviewed,
we now turn to the field on thermal equilibrium states characterized by a finite temperature $T$.

\subsection{Two Parts of the Free Energy}
In quantum field theory, the imaginary-time Mastsubara formalism may be the easiest way to
introduce the influence of temperature to a system.
In this formalism one makes the time coordinate rotate as
$t\rightarrow-\mathrm{i}\tau$ and the Euclidean time $\tau$ is confined to the interval
$\tau$$\in$[0, $\beta$], where $\beta=1/T$. Periodic BC $\varphi(\tau+\beta,
 \textbf{x})=\varphi(\tau,\textbf{x})$ for bosonic field are imposed in the imaginary time coordinate.
The partition function $\mathcal{Z}$ is given by
\begin{equation}\label{eq1}
	\mathcal{Z}=C\int\Diff\varphi\e^{-S_\text{E}[\varphi]},
\end{equation}
where $S_\text{E}[\varphi]$ is the Euclidean action
\begin{equation}\label{eq2}
	S_\text{E}[\varphi]=\frac{1}{2}\int_ 0^\beta\diff\tau\int\diff^D\mathbf{x} \varphi K_\text{E} \varphi,
\end{equation}
with $K_\text{E}=-\square_\text{E}$ and $\square_\text{E}=\frac{\partial^2 }{\partial
\tau^2}+\vartriangle$ being Euclidean wave operator.
Then the Helmholtz free energy can be expressed as
\begin{equation}
	F=-\frac 1 {\beta}\log(\mathcal{Z})=\frac1{2\beta}\mathrm{Tr}\log (K_\text{E}).
\end{equation}

The configuration is still a $p$-dimensional hypercubic cavity with the size $L_1=L_2=\cdots=L_p=L$
and with the sizes of the left $(D-p)$-dimension $L_{p+1},L_{p+2},\cdots,L_D\gg L$ in $(D+1)$-dimensional spacetime.
When the scalar field satisfies periodic, Dirichlet and Neumann BCs,
the Helmholtz free energies have the following expressions, respectively
\begin{equation}
\begin{split}
	F^{(\text{P})}=&\frac1{2\beta}(\prod_{j=p+1}^D\frac{L_j}{2\pi})\sum_{\substack{n_0\in\mathbb{Z}\\\vec{n}\in\mathbb{Z}^p}}\int_{-\infty}^\infty\diff^{D-p}\mathbf{r}\\
	 &\times\log[(\frac{2\pi n_0}{\beta})^2+(\frac{2\pi n_1}{L})^2+\cdots+(\frac{2\pi n_p}{L})^2+\mathbf{r}^2],\\
	\label{HelmP}
\end{split}
\end{equation}
\begin{equation}
\begin{split}
	F^{(\text{D})}=&\frac1{2\beta}(\prod_{j=p+1}^D\frac{L_j}{\pi})\sum_{\substack{n_0\in\mathbb{Z}\\\vec{n}\in\mathbb{N}^p}}\int_{-\infty}^\infty\diff^{D-p}\mathbf{r}\\
	&\times\log[(\frac{2\pi n_0}{\beta})^2+(\frac{\pi n_1}{L})^2+\cdots+(\frac{\pi n_p}{L})^2+\mathbf{r}^2],\\
	\label{HelmD}
\end{split}
\end{equation}
\begin{equation}
\begin{split}
	F^{(\text{N})}=&\frac1{2\beta}(\prod_{j=p+1}^D\frac{L_j}{\pi})\sum_{\substack{n_0\in\mathbb{Z}\\\vec{n}\in(\mathbb{N}\cup\{\vec{0}\})^p}}\int_{-\infty}^\infty\diff^{D-p}\mathbf{r}\\
	&\times\log[(\frac{2\pi n_0}{\beta})^2+(\frac{\pi n_1}{L})^2+\cdots+(\frac{\pi n_p}{L})^2+\mathbf{r}^2].
	\label{HelmN}
\end{split}
\end{equation}

On the regularization of the divergency, Mellin transformation and zeta function technique still come in handy.
The density of the free energy of the periodic case
\begin{equation}
	\begin{split}
	 f^{(\text{P})}\equiv&\frac{F^{(\text{P})}}{\prod_{j=p+1}^DL_j}\\
	 =&-\frac1{2^{D-p}\pi^{\frac{D-p}{2}}\Gamma(\frac{D-p}{2})\beta}\\
	 &\times\sum_{\substack{n_0\in\mathbb{Z}\\\vec{n}\in\mathbb{Z}^p}}\lim_{s\rightarrow0}\frac{\partial}{\partial s}\int_0^\infty\frac{r^{D-p-1}}{\Gamma(s)}\int_0^\infty t^{s-1}\e^{-[(\frac{2\pi n_0}{\beta})^2+\frac{4\pi^2\vec{n}^2}{L^2}+r^2]t}\diff t\diff r.
	\label{densP}
\end{split}
\end{equation}
With the Poisson summation eq.\eqref{poisson} employed on the $n_0$ summation,
eq.\eqref{densP} becomes
\begin{equation}
	 f^{(\text{P})}=-\frac1{2^{D-p+1}\pi^{\frac{D-p}{2}}}\sum_{\substack{m_0\in\mathbb{Z}\\\vec{n}\in\mathbb{Z}^p}}\int_0^\infty t^{-1-\frac{D-p}{2}}\frac1{\sqrt{4\pi t}}\e^{-\frac{m_0^2\beta^2}{4t}}\e^{-\frac{4\vec{n}^2\pi^2}{L^2}t}\diff t.
	\label{densP3}
\end{equation}
The terms $m_0=0$ can be taken out of the summation and $f^{(\text{P})}$ is divided into two parts:
the zero temperature part $\varepsilon_0^{(\text{P})}$
and the temperature-dependent part $f_T^{(\text{P})}$:
\begin{eqnarray}
	\label{e0P}
	\varepsilon_0^{(\text{P})}&=&-\frac1{2L^{D-p+1}}\pi^{\frac{D-p+1}2}\Gamma(-\frac{D-p+1}2)Z_p(p-D-1),\\
	\label{fTP}
	 f_T^{(\text{P})}&=&-\frac2{(\beta L)^{\frac{D-p+1}2}}\sum_{\substack{m_0\in\mathbb{N}\\\vec{n}\in\mathbb{Z}^p}}(\frac{\sqrt{\vec{n}^2}}{m_0})^{\frac{D-p+1}2}K_{\frac{D-p+1}2}(\frac{2\pi m_0\sqrt{\vec{n}^2}\beta}L).
\end{eqnarray}
The former can also be obtained by taking $\beta$ in eq.\eqref{densP} to infinity and turning the summation over $n_0$ into a integral.
As done in previous sections, the regularized zero point energy density is
\begin{equation}
	\varepsilon_0^{(\text{P}),\text{reg.}}=-\frac{\Gamma(\frac{D+1}2)Z_p(D+1)}{2\pi^{\frac{D+1}2}L^{D-p+1}}.
	\label{e0P2}
\end{equation}

As for the temperature-dependent part \eqref{fTP},
it is already finite for a given side length $L$.
This is the very reason that the regularization of this part was neglected in some previous papers.
It is not difficult to find that the free energy density and further the Casimir force density are divergent when $L$ goes large enough,
which contradicts to the fact that the Casimir force should tend to zero with the increase of the side lengths.
So this part of the free energy density is yet to be regularized.
Before that, the divided two parts of the free energy density for the other two BCs are given as follows,
\begin{eqnarray}
	\label{e0DN2}
	\varepsilon_0^{(\text{D/N}),\text{reg}.}&=&-\frac1{2^{D+2}L^{D-p+1}}\sum_{q=0}^{p-1}(\mp1)^q\frac{C_p^q\Gamma(\frac{D-q+1}2)}{\pi^{\frac{D-q+1}2}}Z_{p-q}(D-q+1),\\
	\label{fTDN1}
	f_T^{(\text{D/N})}&=&-\frac2{(2\beta L)^{\frac{D-p+1}2}}\sum_{\substack{m_0\in\mathbb{N}\\\vec{n}\in\mathbb{N}^p/\vec{n}\in(\mathbb{N}\cup\{\vec{0}\})^p}}(\frac{\sqrt{\vec{n}^2}}{m_0})^{\frac{D-p+1}2}K_{\frac{D-p+1}2}(\frac{\pi m_0\sqrt{\vec{n}^2}\beta}L).\quad\quad
\end{eqnarray}

\subsection{The Regularization of the Temperature-Dependent Part}
According to Sect. \ref{equivreg},
it is safe to use ``different'' approach rather than the zeta function technique
to regularize the temperature-dependent part of the free energy.
The Abel-Plana formula eq.\eqref{AbelPlana} is the choice here.
For the cases here, since in eq.\eqref{AbelPlana} the last term is convergent (and will be denoted as $C$ in the following),
the divergent integral is of concern.
With the summation to be regularized denoted as $A\equiv\sum_{n=1}^\infty u(n)$, the regularized $A$ will be
\begin{equation}
\begin{split}
	A^{\text{reg}.}&=\sum_{n=1}^\infty u(n)-\int_0^\infty u(x)\diff x\\
	&=A-\int_0^\infty u(x)\diff x.
	\label{APreg}
\end{split}
\end{equation}

\subsubsection{Periodic BCs}
Denoting
\begin{equation}
	g(z)\equiv-2\sum_{m_0\in\mathbb{N}}(\frac{\sqrt{z}}{m_0\beta L})^{\frac{D-p+1}2}K_{\frac{D-p+1}2}(\frac{2\pi m_0\sqrt{z}\beta}L),
	\label{g}
\end{equation}
then $f_T^{(\text{P})}=\sum_{\vec{n}\in\mathbb{Z}^p}g(\vec{n}^2)$. According to Abel-Plana formula \eqref{AbelPlana},
\begin{equation}
\begin{split}
	 f_T^{(\text{P})}=\sum_{\vec{n}\in\mathbb{Z}^p}g(\vec{n}^2)=&2\sum_{\substack{\vec{n}\in\mathbb{Z}^{p-1}\\k\in\mathbb{N}}}g(\vec{n}^2+k^2)+\sum_{\vec{n}\in\mathbb{Z}^{p-1}}g(\vec{n}^2)\\
	=&2\sum_{\vec{n}\in\mathbb{Z}^{p-1}}\Big[-\frac{1}{2}g(\vec{n}^2)+\int_0^\infty g(\vec{n}^2+x^2)\diff x\Big]+\sum_{\vec{n}\in\mathbb{Z}^{p-1}}g(\vec{n}^2)+C\\
	=&2\sum_{\vec{n}\in\mathbb{Z}^{p-1}}\int_0^\infty g(\vec{n}^2+x^2)\diff x+C\\
	=&2^p\int_0^\infty g(x_1^2+\cdots+x_p^2)\diff^px+C.\\
	\label{fTPg}
\end{split}
\end{equation}
That is
\begin{equation}
\begin{split}
	f_T^{(\mathrm{P})}=&-2^{p+1}\sum_{m_0\in\mathbb{N}}\int_0^\infty(\frac{\sqrt{\vec{x}^2}}{m_0\beta L})^{\frac{D-p+1}{2}}K_{\frac{D-p+1}{2}}(\frac{2m_0\beta\pi\sqrt{\vec{x}^2}}{L})\diff^p\vec{x}+C\\
	=&-\frac{L^p\Gamma(\frac{D+1}{2})\zeta(D+1)}{\beta^{D+1}\pi^{\frac{D+1}{2}}}+C.
	\label{fTP2}
\end{split}
\end{equation}
Now, according to eq.\eqref{APreg} it is clearly seen that to get the regularized result,
the term has to be subtracted from $f_T^{(\text{P})}$ is
\begin{equation}
	-\frac{L^p\Gamma(\frac{D+1}{2})\zeta(D+1)}{\beta^{D+1}\pi^{\frac{D+1}{2}}}.
	\label{subtracttermP}
\end{equation}
Then, the regularized temperature-dependent part of the free energy density is
\begin{equation}
	 f_T^{(\text{P}),\text{reg}.}=f_T^{(\text{P})}+\frac{L^p\Gamma(\frac{D+1}{2})\zeta(D+1)}{\beta^{D+1}\pi^{\frac{D+1}{2}}}.
	\label{fTP3}
\end{equation}
\subsubsection{Dirichlet and Neumann BCs}
For Dirichlet and Neumann BCs, similarly, denoting
\begin{equation}
	h(z)\equiv-2\sum_{m_0\in\mathbb{N}}(\frac{\sqrt{z}}{2\beta Lm_0})^{\frac{D-p+1}{2}}K_{\frac{D-p+1}{2}}(\frac{m_0\pi\beta\sqrt{z}}{L}),
	\label{h}
\end{equation}
then
\begin{equation}
	f_T^{(\text{D})}=\sum_{\vec{n}\in\mathbb{N}^p}h(\vec{n}^2),\quad f_T^{(\text{N})}=\sum_{\vec{n}\in(\mathbb{N}\cup\{\vec{0}\})^p}h(\vec{n}^2).
	\label{fTDNh}
\end{equation}
According to eq.\eqref{AbelPlana}
\begin{equation}
\begin{split}
	 f_T^{(\text{D/N})}=&\sum_{\vec{n}\in\mathbb{N}^p/\vec{n}\in(\mathbb{N}\cup\{\vec{0}\})^p}h(\vec{n}^2)=\sum_{\vec{n}\in\mathbb{N}^{p-1}/\vec{n}\in(\mathbb{N}\cup\{\vec{0}\})^{p-1}}\sum_{k=1/0}^\infty h(\vec{n}^2+k^2)\\
	 =&\sum_{\vec{n}\in\mathbb{N}^{p-1}/\vec{n}\in(\mathbb{N}\cup\{\vec{0}\})^{p-1}}\Big[\mp\frac{1}{2}h(\vec{n}^2)+\int_0^\infty h(\vec{n}^2+x^2)\diff x\Big]+C.\\
	\label{fTDN2}
\end{split}
\end{equation}

Following the procedure in Ref. \refcite{Lin2014} one has
\begin{equation}
\begin{split}
	 f_T^{(\text{D/N})}=&\sum_{q=0}^{p-1}C_p^q(\mp\frac{1}{2})^q\int_0^\infty h(x_1^2+\cdots+x_{p-q}^{2})\diff^{p-q}x+(\mp\frac{1}{2})^ph(\vec{0}^2)+C\\
	=&-\sum_{q=0}^{p-1}(\mp1)^q\frac{C_p^qL^{p-q}\zeta(D-q+1)}{2^q\pi^{\frac{D-q+1}{2}}\beta^{D-q+1}}\Gamma(\frac{D-q+1}{2})\\
	 &-\frac{(\mp1)^p\Gamma(\frac{D-p+1}{2})}{2^p\beta^{D-p+1}\pi^{\frac{D-p+1}{2}}}\zeta(D-p+1)+C.\\
\end{split}
\end{equation}
The related terms contributing to the divergency of the Casimir force,
with which one is dealing in the first place, are
\begin{equation}
	 -\sum_{q=0}^{p-1}(\mp1)^q\frac{C_p^qL^{p-q}\zeta(D-q+1)}{2^q\pi^{\frac{D-q+1}{2}}\beta^{D-q+1}}\Gamma(\frac{D-q+1}{2}),
	\label{subtracttermDN}
\end{equation}
which should be subtracted from the free energy.
So, the regularized temperature-dependent parts of the free energy densities for these two BCs are
\begin{equation}
	 f_T^{(\text{D/N}),\text{reg}.}=f_T^{(\mathrm{D/N})}+\sum_{q=0}^{p-1}(\mp1)^q\frac{C_p^qL^{p-q}\zeta(D-q+1)}{2^q\pi^{\frac{D-q+1}{2}}\beta^{D-q+1}}\Gamma(\frac{D-q+1}{2}).
	\label{fTDN3}
\end{equation}

In the case of periodic BCs,
only one term needs to be subtracted,
namely eq.\eqref{subtracttermP},
while in the cases of Dirichlet and Neumann BCs,
there are $p$ terms as shown in eq.\eqref{subtracttermDN}.
For $D=p=3$, from eq.\eqref{subtracttermP}, one gets
\begin{equation}
	-\frac{L^3\pi^2T^4}{90},
\end{equation}
and from eq.\eqref{subtracttermDN}, one gets
\begin{equation}
	-\frac{L^3\pi^2T^4}{90},\quad\quad\pm\frac{3\zeta(3)L^2T^3}{4\pi},\quad\quad-\frac{L\pi T^2}8,
	\label{subtractterm3}
\end{equation}
where the sign ``$+$'' corresponds to Dirichlet BCs and the sign ``$-$'' to Neumann BCs.
It is obvious that the term proportional to $T^4$ is the blackbody radiation energy restricted in the volume $L^3$,
regardless of the BCs.
For Dirichlet BCs the three terms in \eqref{subtractterm3} are the results obtained in the previous papers\cite{Dowker1978,Nesterenko2003,Vassilevich2003}.
Therefore, it can be said that eqs.\eqref{subtracttermP} and \eqref{subtracttermDN} are the general results of the subtraction
to get the physical Casimir free energy density for $p$-dimensional hypercube in $(D+1)$-dimensional spacetime.

The explicit expression of the terms to be subtracted here
are indeed of order equal to or more than the square of the temperature as suggested in Refs. \refcite{Bezerra2011,Geyer2008}.
As emphasized before, when the side length tends to infinity
the unregularized temperature-dependent part of the free energy density is divergent,
and now one can see the terms to be subtracted are exactly proportional to the powers of the side length.

At this point, we write out the expressions of the physical free energy for all three kinds of BCs:
\begin{equation}
\begin{split}
	f^{\text{(P)},\text{Phys.}}=&\varepsilon_0^{(\text{P}),\text{reg}.}+f_T^{(\text{P}),\text{reg}.}\\
	 =&-\frac{\Gamma(\frac{D+1}{2})Z_p(D+1)}{2\pi^{\frac{D+1}{2}}L^{D-p+1}}-2\sum_{\substack{m_0\in\mathbb{N}\\\vec{n}\in\mathbb{Z}^p}}(\frac{\sqrt{\vec{n}^2}}{m_0\beta L})^{\frac{D-p+1}{2}}K_{\frac{D-p+1}{2}}(\frac{2m_0\beta\pi\sqrt{\vec{n}^2}}{L})\\
	 &+\frac{L^p\zeta(D+1)\Gamma(\frac{D+1}{2})}{\pi^{\frac{D+1}{2}}\beta^{D+1}},
	\label{fPhysP}
\end{split}
\end{equation}
and
\begin{equation}
\begin{split}
	f^{(\text{D/N}),\text{Phys.}}=&\varepsilon_0^{(\text{D/N}),\text{reg}.}+f_T^{(\text{D/N}),\text{reg}.}\\
	 =&-\frac{1}{2^{D+2}L^{D-p+1}}\sum_{q=0}^{p-1}\frac{C_p^q(\mp1)^q\Gamma(\frac{D-q+1}{2})}{\pi^{\frac{D-q+1}{2}}}Z_{p-q}(D-q+1)\\
	 &-2\sum_{m_0\in\mathbb{N}}\sum_{\vec{n}\in\mathbb{N}^p/\vec{n}\in(\mathbb{N}\cup\{\vec{0}\})^p}(\frac{\sqrt{\vec{n}^2}}{2\beta Lm_0})^{\frac{D-p+1}{2}}K_{\frac{D-p+1}{2}}(\frac{m_0\pi\beta\sqrt{\vec{n}^2}}{L})\\
	 &+\sum_{q=0}^{p-1}(\mp1)^q\frac{C_p^qL^{p-q}\zeta(D-q+1)}{2^q\pi^{\frac{D-q+1}{2}}\beta^{D-q+1}}\Gamma(\frac{D-q+1}{2}).
	\label{fPhysDN}
\end{split}
\end{equation}

\subsection{Alternative Expressions of the Casimir Free Energy}
The terms like $K_\nu(\alpha\frac{\beta}L)$ in eqs.\eqref{fTP} and \eqref{fTDN3},
which describe the cases of low temperature or small separations better since $K_\nu(z)$ converges fast for large $z$,
come from the employment of the Poisson summation formula over the $n_0$ summation.
One can also employ the Poisson summation formula over the $\vec{n}$ summations,
which will results in terms like $K_{\nu'}(\alpha'\frac L{\beta})$ that describe the cases of high temperature or large separations better.
The two kinds of results are equivalent for the same case,
and should be called the low temperature and high temperature expansions respectively
for the only difference lies in the converging rapidness in different temperature regimes.
Through the similar procedure,
one can get the following high temperature expansions of the free energy densities:
\begin{equation}
\begin{split}
	f'^{\text{(P)}}=&-\frac{\Gamma(\frac{D}{2})}{2\beta L^{D-p}\pi^{\frac{D}{2}}}Z_p(D)-\frac{2L^p}{\beta}\sum_{\substack{n_0\in\mathbb{N}\\\vec{m}\in\mathbb{Z}^p\setminus\{\vec{0}\}}}(\frac{n_0}{\beta L\sqrt{\vec{m}^2}})^{\frac{D}{2}}K_{\frac{D}{2}}(\frac{2n_0\pi L}{\beta}\sqrt{\vec{m}^2})\\
	&-\frac{L^p\zeta(D+1)\Gamma(\frac{D+1}{2})}{\pi^{\frac{D+1}{2}}\beta^{D+1}},\\
	\label{f'TP1}
\end{split}
\end{equation}
\begin{equation}
\begin{split}
	 f'^{(\text{D/N})}=&-\frac{(\mp1)^p\Gamma(\frac{D-p+1}{2})}{2^p\beta^{D-p+1}\pi^{\frac{D-p+1}{2}}}\zeta(D-p+1)\\
	 &-\frac{1}{2^{D+1}\beta L^{D-p}}\sum_{q=0}^{p-1}\frac{(\mp1)^qC_p^q\Gamma(\frac{D-q}{2})}{\pi^{\frac{D-q}{2}}}Z_{p-q}(D-q)\\
	 &-\frac{1}{2^{D-1}\beta}\sum_{q=0}^{p-1}C_p^q(\mp1)^qL^{p-q}\sum_{\substack{n_0\in\mathbb{N}\\\vec{m}\in\mathbb{Z}^{p-q}\setminus\{\vec{0}\}}}(\frac{2n_0}{L\beta\sqrt{\vec{m}^2}})^{\frac{D-q}{2}}K_{\frac{D-q}{2}}(\frac{4\pi n_0L\sqrt{\vec{m}^2}}{\beta})\\
	 &-\sum_{q=0}^{p-1}(\mp1)^q\frac{C_p^qL^{p-q}\zeta(D-q+1)}{2^q\pi^{\frac{D-q+1}{2}}\beta^{D-q+1}}\Gamma(\frac{D-q+1}{2}).
	\label{f'TDN1}
\end{split}
\end{equation}

Since the two expansions of the free energy densities in low and high temperature regimes are equivalent,
finite physical results should be obtained from both expansions when $L\rightarrow\infty$.
Now, in high temperature expansions \eqref{f'TP1} and \eqref{f'TDN1},
it is easy to see that the divergent terms as $L\rightarrow\infty$ are the last terms of each equation.
So these terms, which coincide with eqs.\eqref{subtracttermP} and \eqref{subtracttermDN}, have to be removed.
Then, the physical free energy densities in high temperature regime are expressed as
\begin{equation}
\begin{split}
	f'^{\text{(P)},\text{Phys}.}=&-\frac{\Gamma(\frac{D}{2})}{2\beta L^{D-p}\pi^{\frac{D}{2}}}Z_p(D)\\
	&-\frac{2L^p}{\beta^{\frac{D}{2}+1}L^{\frac{D}{2}}}\sum_{\substack{n_0\in\mathbb{N}\\\vec{m}\in\mathbb{Z}^p\setminus\{\vec{0}\}}}n_0^{\frac{D}{2}}(\vec{m}^2)^{-\frac{D}{4}}K_{\frac{D}{2}}(\frac{2n_0\pi L}{\beta}\sqrt{\vec{m}^2}),\\
	\label{f'PhysP}
\end{split}
\end{equation}
and
\begin{equation}
\begin{split}
	 f'^{(\text{D/N}),\text{Phys}.}=&-\frac{(\mp1)^p\Gamma(\frac{D-p+1}{2})}{2^p\beta^{D-p+1}\pi^{\frac{D-p+1}{2}}}\zeta(D-p+1)\\
	 &-\frac{1}{2^{D+1}\beta L^{D-p}}\sum_{q=0}^{p-1}\frac{(\mp1)^qC_p^q\Gamma(\frac{D-q}{2})}{\pi^{\frac{D-q}{2}}}Z_{p-q}(D-q)\\
	 &-\Big[\frac{1}{2^{D-1}\beta}\sum_{q=0}^{p-1}C_p^q(\mp1)^qL^{p-q}\\
	 &\times\sum_{\substack{n_0\in\mathbb{N}\\\vec{m}\in\mathbb{Z}^{p-q}\setminus\{\vec{0}\}}}(\frac{2n_0}{L\beta\sqrt{\vec{m}^2}})^{\frac{D-q}{2}}K_{\frac{D-q}{2}}(\frac{4\pi n_0L\sqrt{\vec{m}^2}}{\beta})\Big].
	\label{f'PhysDN}
\end{split}
\end{equation}

\subsection{The Closed Case of $D=p$}
The closed case of $D=p$ has been investigated by
Ambj{\o}rn and Wolfram\cite{Ambjorn1983} and Lim and Teo\cite{Lim2007} for periodic BCs.
By making some modifications of the results of $D>p$ case,
we review both the low and high temperature expansions for all three BCs.
\subsubsection{Low Temperature Expansion}
From eqs.\eqref{fPhysP} and \eqref{fPhysDN}, when $\vec{n}\in\{\vec{0}\}$,
the second terms becomes (for periodic and Neumann BCs but not for Dirichlet)
\begin{equation}
	-\frac{\Gamma(\frac{D-p+1}2)\zeta(D-p+1)}{\pi^{\frac{D-p+1}2}\beta^{D-p+1}},
	\label{subtracttermd=p}
\end{equation}
which is divergent for $D=p$.
However, Ambj{\o}rn and Wolfram\cite{Ambjorn1983} have argued physically that this term is the free Bose gas result
and should not appear in the result of physical free energy of $D=p$ case.
Therefore, with this term excluded,
the physical free energies for a closed $D=p$ cavity in low temperature expansion are
\begin{equation}
\begin{split}
	 f^{\text{(P)},\text{Phys.}}_{D=p}=&-\frac{\Gamma(\frac{D+1}{2})Z_D(D+1)}{2\pi^{\frac{D+1}{2}}L}-2\sum_{\substack{m_0\in\mathbb{N}\\\vec{n}\in\mathbb{Z}^D\setminus\{\vec{0}\}}}(\frac{\sqrt{\vec{n}^2}}{m_0\beta L})^{\frac{1}{2}}K_{\frac{1}{2}}(\frac{2m_0\beta\pi\sqrt{\vec{n}^2}}{L})\\
	 &+\frac{L^p\zeta(D+1)\Gamma(\frac{D+1}{2})}{\pi^{\frac{D+1}{2}}\beta^{D+1}},
	\label{fPhysPd=p}
\end{split}
\end{equation}
and
\begin{equation}
\begin{split}
	 f^{(\text{D/N}),\text{Phys.}}_{D=p}=&-\frac{1}{2^{D+2}L}\sum_{q=0}^{D-1}\frac{C_D^q(\mp1)^q\Gamma(\frac{D-q+1}{2})}{\pi^{\frac{D-q+1}{2}}}Z_{D-q}(D-q+1)\\
	 &-2\sum_{m_0\in\mathbb{N}}\sum_{\vec{n}\in\mathbb{N}^D/\vec{n}\in(\mathbb{N}\cup\{\vec{0}\})^D}{'}(\frac{\sqrt{\vec{n}^2}}{2\beta Lm_0})^{\frac{1}{2}}K_{\frac{1}{2}}(\frac{m_0\pi\beta\sqrt{\vec{n}^2}}{L})\\
	 &+\sum_{q=0}^{D-1}(\mp1)^q\frac{C_D^qL^{p-q}\zeta(D-q+1)}{2^q\pi^{\frac{D-q+1}{2}}\beta^{D-q+1}}\Gamma(\frac{D-q+1}{2}).
	\label{fPhysDNd=p}
\end{split}
\end{equation}
\subsubsection{High Temperature Expansion}
Eq.\eqref{subtracttermd=p} is still to be removed in the high temperature expansion
to obtain the physical free energies for $D=p$ case.
However, it is not that straightforward to get to the results as in low temperature regime.
From eq.\eqref{f'PhysP} one can see that the first term is also divergent for $D=p$.
So, we have to deal with two divergent terms now.

Recall that the Epstein zeta function can be written as eq.\eqref{CS2}.
Replacing the argument and it can be rewritten as
\begin{equation}
	Z_p(s)=\frac2{\Gamma(\frac s2)}\sum_{j=0}^{p-1}\pi^{\frac j2}\Gamma(\frac{s-j}2)\zeta(s-j)+\frac{4\pi^{\frac s2}}{\Gamma(\frac s2)}\sum_{j=1}^{p-1}\sum_{\substack{m\in\mathbb{N}\\\vec{k}\in\mathbb{Z}^j\setminus\{\vec{0}\}}}(\frac{\sqrt{\vec{k}^2}}m)^{\frac{s-j}2}K_{\frac{s-j}2}(2\pi m\sqrt{\vec{k}^2}).
	\label{epsteinexpansion}
\end{equation}
When $s=p$, the divergency lies only in the $j=p-1$ term of the first part.
So, the first term of eq.\eqref{f'PhysP} can be expressed as
\begin{equation}
\begin{split}
	-\frac{\Gamma(\frac D2)Z_p(D)}{2\beta L^{D-p}\pi^{\frac D2}}=&-\frac{\Gamma(\frac{D-p+1}2)\zeta(D-p+1)}{\beta L^{D-p}\pi^{\frac{D-p+1}2}}-\frac1{\beta L^{D-p}}\sum_{j=0}^{p-2}\frac{\Gamma(\frac{D-j}2)\zeta(D-j)}{\pi^{\frac{D-j}2}}\\
	&-\frac2{\beta L^{D-p}}\sum_{j=1}^{p-1}\sum_{\substack{m\in\mathbb{N}\\\vec{k}\in\mathbb{Z}^j\setminus\{\vec{0}\}}}(\frac{\sqrt{\vec{k}^2}}m)^{\frac{D-j}2}K_{\frac{D-j}2}(2\pi m\sqrt{\vec{k}^2}),
	\label{divergencyP}
\end{split}
\end{equation}
where the first term of the RHS is divergent and the rest are convergent when $D\rightarrow p$.
Now, together with eq.\eqref{subtracttermd=p}, the divergency can be expressed as
\begin{equation}
	-\lim_{D\rightarrow p}\frac{\Gamma(\frac{D-p+1}2)\zeta(D-p+1)}{\pi^{\frac{D-p+1}2}\beta}\left[\frac1{L^{D-p}}-\frac1{\beta^{D-p}}\right]=-\frac1\beta\log(\frac{\beta}L).
	\label{logP}
\end{equation}
Collecting all the pieces,
the free energy for periodic BCs of $D=p$ case in high temperature regime is
\begin{equation}
\begin{split}
	 f^{\text{(P)},\text{Phys}.}_{D=p}=&-\sum_{j=0}^{D-2}\frac{\Gamma(\frac{D-j}{2})\zeta(D-j)}{\beta\pi^{\frac{D-j}{2}}}-\frac{2L^{\frac{D}{2}}}{\beta^{\frac{D}{2}+1}}\sum_{\substack{n_0\in\mathbb{N}\\\vec{m}\in\mathbb{Z}^D\setminus\{\vec{0}\}}}(\frac{n_0}{\sqrt{\vec{m}^2}})^{\frac D2}K_{\frac{D}{2}}(\frac{2n_0\pi L}{\beta}\sqrt{\vec{m}^2})\\
	 &-\frac{2}{\beta}\sum_{j=1}^{D-1}\sum_{\substack{m\in\mathbb{N}\\\vec{k}\in\mathbb{Z}^j\setminus\{\vec{0}\}}}(\frac{\sqrt{\vec{k}^2}}{m})^{\frac{D-j}{2}}K_{\frac{D-j}{2}}(2\pi m\sqrt{\vec{k}^2})-\frac{1}{\beta}\ln\frac{\beta}{L}.
	\label{f'PhysPd=p}
\end{split}
\end{equation}

For Dirichlet and Neumann BCs, as $D\rightarrow p$,
the first two terms in eqs.\eqref{f'PhysDN} are both divergent.
Through the procedure similar to eqs.\eqref{divergencyP} to \eqref{f'PhysPd=p},
the physical free energy for $D=p$ case in Dirichlet and Neumann BCs is given as
\begin{equation}
\begin{split}
	 f^{(\text{D/N}),\text{Phys}.}_{D=p}=&-\frac{1}{2^{D-1}\beta}\sum_{q=0}^{D-1}C_D^q(\mp1)^q\sum_{\substack{n_0\in\mathbb{N}\\\vec{m}\in\mathbb{Z}^{D-q}\setminus\{\vec{0}\}}}(\frac{2n_0L}{\beta\sqrt{\vec{m}^2}})^{\frac{D-q}{2}}K_{\frac{D-q}{2}}(\frac{4\pi n_0L\sqrt{\vec{m}^2}}{\beta})\\
	 &-\frac1{2^D\beta}\sum_{q=0}^{D-2}\sum_{j=0}^{D-q-2}\frac{(\mp1)^qC_D^q}{\pi^{\frac{D-q-j}{2}}}\Gamma(\frac{D-q-j}{2})\zeta(D-q-j)\\
	 &-\frac1{2^{D-1}\beta}\sum_{q=0}^{D-2}\sum_{j=1}^{D-q-1}\sum_{\substack{m\in\mathbb{N}\\\vec{k}\in\mathbb{Z}^j\setminus\{\vec{0}\}}}(\mp1)^qC_D^q(\frac{\sqrt{\vec{k}^2}}{m})^{\frac{D-q-j}{2}}K_{\frac{D-q-j}{2}}(2m\pi\sqrt{\vec{k}^2})\\
	 &+A,
	\label{f'PhysDNd=p}
\end{split}
\end{equation}
with
\begin{equation}
	A=
	\begin{cases}
		-\frac{(-1)^D}{2^D\beta}\log(\frac{2L}{\beta})\quad\quad&\text{for Dirichlet BCs};\\
		-\frac{1-2^D}{2^D\beta}\log(\frac{2L}{\beta})\quad\quad&\text{for Neumann BCs}.
	\end{cases}
	\label{logDN}
\end{equation}

The finite temperature cases of all three BCs of $D\ge p$ has been reviewed till now,
expressed in eqs.\eqref{fPhysP}, \eqref{fPhysDN}, \eqref{f'PhysP}, \eqref{f'PhysDN},
\eqref{fPhysPd=p}, \eqref{fPhysDNd=p}, and \eqref{f'PhysPd=p} - \eqref{logDN},
where it is easy to find that
the absolute value of the physical free energy (density) in every case is an increasing function of temperature.
More easy observations of these results are their asymptotic behaviors in high temperature limits,
which are listed in Table \ref{asymptoticf}.
In the table, only the term of the highest power of $\frac1{\beta}$ is retained.
The high temperature limits of the free energy (density) for Dirichlet BCs with even $p$ and for Neumann and periodic BCs both with $D>p$ are negative
whereas they are positive for Dirichlet BCs with odd $p$ and for Neumann and periodic BCs both with $D=p$.
Standard units are also restored in the table,
and it can be seen that with the definition of effective temperature,
$k_\text{B}T_\text{eff}\equiv\hbar c/L$ for periodic BCs, and $2k_\text{B}T_\text{eff}\equiv\hbar c/L$ for the other two BCs\cite{mostepanenko1997},
the high temperature limits of the free energy (density) in all the cases are proportional to $k_\text{B}T$ and do not depend on the Planck constant,
i.e., the classical limits are achieved.
\begin{table}[!htbp]
\renewcommand{\arraystretch}{2.5}
\tbl{High temperature limits of the free energy (density)}
{\begin{tabular}{c|c|c|c}
	\hline
	\hline
	\multicolumn{2}{c|}{ }			&Natural Units	&Standard Units\\
	\hline
	\multirow{2}{*}{periodic}	&$D>p$	&$-\frac{\Gamma(\frac{D}{2})Z_p(D)}{2\pi^{\frac{D}{2}}\beta L^{D-p}}$&$-k_\text{B}T\frac{\Gamma(\frac{D}{2})Z_p(D)}{2\pi^{\frac{D}{2}} L^{D-p}}$\\[2mm]
	\cline{2-4}
	&$D=p$	&$-\frac{1}{\beta}\log(\frac{\beta}{L})$&$k_\text{B}T\log(\frac T{T_\text{eff}})$\\[2mm]
	\hline
	\multirow{2}{*}{Neumann}	&$D>p$	 &$-\frac{\Gamma(\frac{D-p+1}{2})\zeta(D-p+1)}{2^p\beta^{D-p+1}\pi^{\frac{D-p+1}{2}}}$&$-k_\text{B}T\frac{\Gamma(\frac{D-p+1}{2})\zeta(D-p+1)}{2^DL^{D-p}\pi^{\frac{D-p+1}{2}}}(\frac T{T_\text{eff}})^{D-p}$\\[2mm]
	\cline{2-4}
					&$D=p$	 &$-\frac{1-2^D}{2^D\beta}\log(\frac{2L}{\beta})$&$k_\text{B}T\frac{2^D-1}{2^D}\log(\frac{T}{T_\text{eff}})$\\[2mm]
	\hline
	\multirow{2}{*}{Dirichlet}	&$D>p$	 &$-\frac{(-1)^p\Gamma(\frac{D-p+1}{2})\zeta(D-p+1)}{2^p\beta^{D-p+1}\pi^{\frac{D-p+1}{2}}}$&$-k_\text{B}T\frac{(-1)^p\Gamma(\frac{D-p+1}{2})\zeta(D-p+1)}{2^DL^{D-p}\pi^{\frac{D-p+1}{2}}}(\frac T{T_\text{eff}})^{D-p}$\\[2mm]
	\cline{2-4}
					&$D=p$	&$-\frac{(-1)^D}{2^D\beta}\log(\frac{2L}{\beta})$	 &$-k_\text{B}T(-\frac12)^D\log(\frac{T}{T_\text{eff}})$\\[2mm]
	\hline
\end{tabular}
\label{asymptoticf}}
\end{table}

\begin{figure}[htp!]
\centering
\subfigure[]{
\label{p2d5DenergyT}
\includegraphics[width=0.3\textwidth]{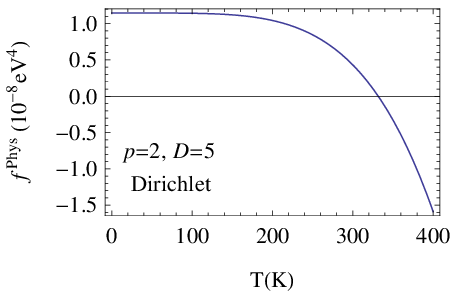}}
\subfigure[]{
\label{p2d6DenergyT}
\includegraphics[width=0.3\textwidth]{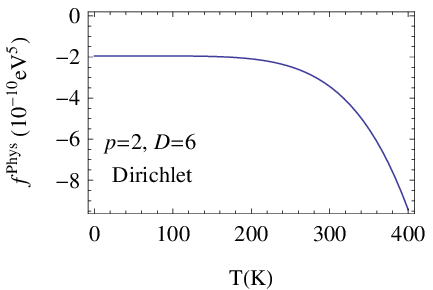}}
\subfigure[]{
\label{p3d5DenergyT}
\includegraphics[width=0.3\textwidth]{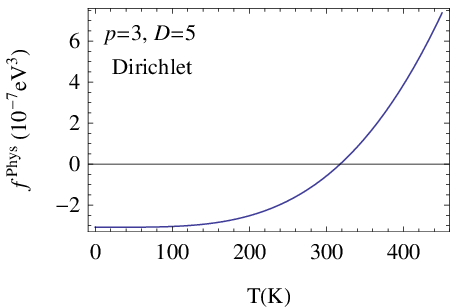}}
\caption{The free energy density as a function of temperature at the side length $L=10\mathrm{eV}^{-1}$:
\textbf{(a)} for Dirichlet BCs with $p=2$, $D=5$;
\textbf{(b)} for Dirichlet BCs with $p=2$, $D=6$;
\textbf{(c)} for Dirichlet BCs with $p=3$, $D=5$.
Note that for $p=2$, $D_\text{crit}=6$ is the critical dimensionality of space
for the zero point energy density given in eq.\eqref{e0DN2}
to turn from positive to negative with $D$ increasing\cite{Caruso1991}.}
\label{fofT}
\end{figure}

\begin{figure}[htp!]
\centering
\subfigure[]{
\label{p2d2DenergyL}
\includegraphics[width=0.35\textwidth]{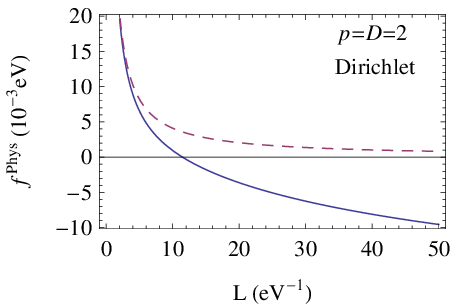}}
\subfigure[]{
\label{p3d3DenergyL}
\includegraphics[width=0.35\textwidth]{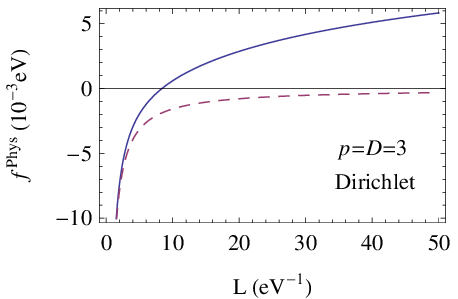}}
\caption{The free energy as a function of the side length for $D=p$ cases.
The solid lines are the free energies of $T=300K$ and the dashed lines are the results of $T=0K$.
\textbf{(a)} is the cases of $D=p=$even of Dirichlet BCs specifying $D=p=2$.
\textbf{(b)} is the cases $D=p=$odd of Dirichlet or any $D=p$ cases of Neumann and periodic BCs, specifying $D=p=3$ in Dirichlet.}
\label{fofLd=p}
\end{figure}

We have collected some numerical analysis of the behaviors of the free energy from Ref. \refcite{Lin2014}.
In Fig. \ref{fofT}, it is shown for Dirichlet BCs the free energy density as a function of temperature at the side length $L=10\mathrm{eV}^{-1}$
with typical dimensions.
And in Fig. \ref{fofLd=p}, it is shown the free energy as a function of the side length for $D=p$ cases,
where the logarithmic behaviors are illustrated,
which although is not seen analytically in the expression in Ref. \refcite{Geyer2008},
is shown in its numerical illustration.
More numerical analyses of the behaviors of the energy can be found in Ref. \refcite{Lin2014}.

\section{Casimir Effect with Helix BCs}
\label{helix}
\setcounter{equation}{0}
The Casimir effect arises not only in the presence of material boundaries,
but also in spaces with nontrivial topology.
For example, a flat manifold with the topology of $S^1$
causes the periodic condition $\phi(t,0)=\phi(t,C)$.

Under the so-called helix boundary condition for a scalar field in 2+1 dimension is defined as
\begin{equation}\label{d2helix}
   \phi(t, x+a, z)= \phi(t,x,z+h)\,,
\end{equation}
where $h$ is regarded as the pitch of the helix,
and this condition is called the helix BC.
One can see that it would return to the cylindrical boundary conditions when $h$ vanishes,
but for $h\neq 0$, the whole system (the spring) does not have the cylindrical symmetry.
In Sect. \ref{helixscalar},  we shall review this kind of boundary coniditions from the lattices aspect in ($D+1$) dimensions.

The Casimir force on the $x$ direction of the helix  can be obtained by using the $\zeta$ function regularization:
\begin{equation}\label{force1 a}
	\mathcal{F}^{(a)} = -\frac{\partial \mathcal{E}^\text{reg.}(a,h)}{\partial a} = - \frac{3\zeta(3)}{2\pi a^4} \bigg(1+r^2\bigg)^{-5/2},
\end{equation}
which is always an attractive force and the magnitude of the force monotonously decreases with the increasing of the ratio $r$.
Once $r$ becomes large enough, the force can be neglected.
While, the Casimir force on the $z$ direction is
\begin{equation}\label{force1 h}
	\mathcal{F}^{(h)} = -\frac{\partial \mathcal{E}^\text{reg.}(a,h)}{\partial h} = - \frac{3\zeta(3)}{2\pi a^4} \frac{r}{(1+r^2)^{5/2}},
    \,.
\end{equation}
which has a maximum magnitude at $r \equiv h/a =0.5$.
When $r<0.5$, the magnitude of the force increases with the increasing of $r$ until $r=0.5$.
The Casimir force is almost linearly depending on $r$ when $r\ll1$,
which is just like the force on a spring complying with the Hooke's law.
However, in this case, the force  originates from the quantum effect, namely, the Casimir effect.
And then, it is called \textit{quantum spring}.
One shall see that the \textit{quantum spring} can exist in any ($D+1$)-dimensional spacetime.

There are also another interesting non-Euclidean topology BC inspired by Nanotubes,
which could be given by introducing an arbitrary phase difference between
$\phi(t, \mathbf{x}+ \mathbf{a}) $ and $ \phi(t, \mathbf{x})$, namely,
 \begin{equation} \label{bpcond}
	\phi(t, \mathbf{x}+ \mathbf{a}) = e^{i2\pi \theta} \phi(t, \mathbf{x})  \,,
\end{equation}
where the phase angle takes the value between $0\leq\theta\leq1$.
Clearly, it will reduce to the (anti-) periodic BC when $\theta$ takes an (half-) integer value.

Generally, the phase could be any values besides $-1$ and $1$ in the complex plane.
For instance, when one considers the Casimir effect in nanotubes or nanoloopes for a quantum field,
$\theta = 0$ corresponds to metallic nanotubes,
while $\theta = \pm 2\pi/3$ corresponds to semiconductor nanotubes.
So, it is more reasonable and interesting to take the this kind of ``quasi-periodic" BC (\ref{bpcond}) for the scalar field.
In Ref. \refcite{FENG2014}, the authors have studied the Casimire effect with the BC (\ref{bpcond}),
and they found that an attractive or repulsive Casimir force could be arised
depending on the values of the phase angle.
Especially, the Casimir effect disappears when the phase angle takes a particular value.
They have also investigated the high dimensional spacetime cases.

In $3+1$ dimensional spacetime that are most interested, the Casimir force could be obtained as
\begin{equation}
	\mathcal{F}_{3}^{(a)} = \frac{ 4\pi^{2 } }    { 3a^{5}} \bigg(-\frac{1}{30} +\theta^{2}  -2\theta^{3} +\theta^{4}  \bigg) \,.
\end{equation}
Cleary, the force is attractive or repulsive depending on the values of $\theta$,
and it could be even vanished when
\begin{equation}
\theta = \frac{1}{2} \pm \frac{1}{2}\sqrt{1-\frac{2\sqrt{30}}{15}}\,.
\end{equation}
This phenomena is so intersting that it worth further studing,
especially by combining the ``quasi-periodic" BC (\ref{bpcond}) with the helix topological condition like Eq.(\ref{d2helix}):
\begin{equation}\label{d2helixa}
   \phi(t, x+a, z)=  e^{i2\pi \theta}\phi(t,x,z+h)\,.
\end{equation}

In the rest of this section, we will review the Casimir effect with helix BCs in some cases.

\subsection{Scalar Casimir Effect }
\label{helixscalar}
\subsubsection{Topology of the Flat ($D$+1)-Dimensional Spacetime}
Prior to the discussion of more complicated topology in the flat spacetime,
it is beneficial to review the idea of lattices.
A lattice $\Lambda$ is defined as a set of points in a flat
($D+1$)-dimensional spacetime $\mathcal{M}^{D+1}$, of the form
\begin{equation}\label{lattice}
    \Lambda = \left\{ ~ \sum_{i=0}^{D} n_i \mathbf{e}_i ~|~ n_i \in \mathbb{Z} ~\right\} \,,
\end{equation}
where $\{\mathbf{e}_i\}$ is a set of basis vectors of $\mathcal{M}^{D+1}$.
In terms of the components $v^i$ of vectors $\mathbf{V} \in \mathcal{M}^{D+1} $,
the inner products is defined as
\begin{equation}\label{inner prod}
    \mathbf{V} \cdot \mathbf{W} = \epsilon(a)v^iw^j\delta_{ij} \,,
\end{equation}
with $\epsilon(a)=1$ for $i=0$, $\epsilon(a)=-1$ for otherwise.
In the $x^1-x^2$ plane, the sublattice
$\Lambda''\subset\Lambda'\subset\Lambda$ are
\begin{equation}\label{sub1}
   \Lambda' = \left\{ ~  n_1 \mathbf{e}_1 + n_2 \mathbf{e}_2 ~|~ n_{1,2} \in \mathbb{Z} ~\right\} \,,
\end{equation}
and
\begin{equation}\label{sub2}
   \Lambda'' = \left\{ ~  n(\mathbf{e}_1 + \mathbf{e}_2) ~|~ n \in \mathbb{Z} ~\right\} \,.
\end{equation}

The unit cylinder-cell is the set of points
\begin{eqnarray}
 \nonumber
   U_c &=& \bigg\{\mathbf{X} = \sum_{i=0}^{D}x^i \mathbf{e}_i ~|~ 0\leq x^1 < a,
 -h\leq x^2 < 0 , \\ && -\infty <x^0<\infty, -\frac{L}{2} \leq x^T\leq \frac{L}{2}\bigg\} \,,\label{cell}
\end{eqnarray}
where $T = 3,\cdots, D$.
When $L\rightarrow\infty$, it contains precisely one lattice point (i.e. $\mathbb{X} = 0$),
and any vector $\mathbb{V}$ has precisely one "image" in the unit cylinder-cell,
obtained by adding a sublattice vector to it.

For the scalar Casimir effect,
a topology of the flat ($D+1$)-dimensional spacetime:
$U_c\equiv U_c +\mathbf{u}, \mathbf{u} \in \Lambda''$ is considered.
This topology causes the helix BCs for a massless or massive scalar field
\begin{equation}\label{helix boundary condition}
   \phi(t, x^1 + a, x^2, x^T) =  \phi(t, x^1 , x^2+h, x^T) \,,
\end{equation}
where, if $a=0$ or $h=0$, it returns to the periodic BCs.

\subsubsection{Massless Scalar Field}
Under the BCs eq.\eqref{helix boundary condition} in ($D+1$)-dimensional flat spacetime,
the eigenfunctions of the massless scalar field satisfying the Klein-Gordon equation are
\begin{equation}\label{modes}
    \phi_{n}(t, x^\alpha, x^T)= \mathcal{N} e^{-i\omega_nt+ik_x x+ik_z z + ik_Tx^T }\,,
\end{equation}
where $\alpha=1,2;\:T=3,\cdots,D$, $\mathcal{N}$ is a normalization factor and $x^1=x, x^2=z$, and
\begin{equation}\label{energyf}
    \omega_n^2 = k_{T}^2 + k_x^2 + \left( -\frac{2\pi n}{h}+\frac{k_x}{h}a \right)^2 = k_{T}^2 + k_z^2 + \left( \frac{2\pi n}{a}+\frac{k_z}{a}h
    \right)^2 \,.
\end{equation}
Here, $k_x$ and $k_z$ satisfy
\begin{equation}\label{kxkzf}
    a k_x - hk_z = 2n\pi\,, (n=0,\pm1,\pm2,\cdots) \,.
\end{equation}
Thus the energy is given as
\begin{eqnarray}
	\mathcal{E}_D =\frac1{2a} \int \frac{d^{D-1}k}{(2\pi)^{D-1}} \sum_{n=-\infty}^{\infty} \sqrt{k_T^2 + k_z^2 + \left( \frac{2\pi n}{a}+\frac{k_z}{a}h  \right)^2  }
	\label{totenergy}
\end{eqnarray}
where it is assumed that $a\neq 0$ without losing generalities.
Eq.\eqref{totenergy} can be rewritten as
\begin{equation}
	\begin{split}
		\mathcal{E}_D =&\frac{1}{2 a \sqrt{\gamma}} \int \frac{d^{D-1}u}{(2\pi)^{D-1}} \sum_{n=-\infty}^{\infty} \sqrt{u^2 + \left( \frac{2\pi n}{a \gamma}\right)^2  }\\
		=&-\frac {\pi^{\frac D 2}}{a^{D+1}\gamma^{\frac {D+1}{2}}}\Gamma \left (-\frac D 2 \right )\zeta(-D),
	\end{split}
	\label{re-totenergy}
\end{equation}
with $\gamma \equiv 1+ \frac {h^2}{a^2}$.
Eq.\eqref{re-totenergy} can be regularized utilizing eq.\eqref{RiemannZetaReflect}.
For $D=2j+1$,
\begin{equation}
	\mathcal{E}^\text{reg.}_{2j+1}=-\frac{(2\pi)^{j+1}|B_{2j+2}|}{(2j+1)!!(2j+2)(a^2+h^2)^{j+1}},
\end{equation}
where $j=1,2,...$ and the Bernoulli numbers are $B_2=\frac16,B_4=-\frac1{30},B_6=\frac1{42},
B_8=-\frac1{30},B_{10}=\frac5{66},B_{12}=-\frac{691}{2730},B_{14}=\frac76,B_{16}=-\frac{3617}{510},\cdots$.
For $D=2j$,
\begin{equation}
	\mathcal{E}^\text{reg.}_{2j}=-\frac{(2j-1)!!\zeta(2j+1)}{(2\pi)^j(a^2+h^2)^{j+\frac 1 2}}
\end{equation}
The symmetry of $a\leftrightarrow h$ is obvious in both cases.
It is worth noting that the Casimir energy has different expressions for the odd and even space dimensions.

The Casimir force on the $x$ direction is
\begin{equation}
	\mathcal{F}^{(a)}=-\frac {\partial\mathcal{E}_D^\text{reg.}}{\partial a}
\end{equation}
In the case of odd-dimensional space, the Casimir force is calculated
\begin{equation}
	\mathcal{F}^{(a)}_{2j+1}=-\frac{(2\pi)^{j+1}|B_{2j+2}|a}{(2j+1)!!(a^2+h^2)^{j+2}},
\end{equation}
which has a maximum value of magnitude
\begin{equation}
	\mathcal{F}_{2j+1}^{(a),\text{max}}=-\frac{(2\pi)^{j+1}|B_{2j+2}|}{(2j+1)!!h^{2j+3}}\sqrt{\frac{(2j+3)^{2j+3}}{(2j+4)^{2j+4}}}
\end{equation}
at $a=\frac {h}{\sqrt{2j+3}}$.
In the case of even-dimensional space, the Casimir force is
\begin{equation}
	\mathcal{F}_{2j}^{(a)}=-\frac{2(j+\frac 1 2)(2j-1)!!\zeta(2j+1)a}{(2\pi)^j(a^2+h^2)^{j+\frac 3 2}},
\end{equation}
and the maximum value of force magnitude
\begin{equation}
	\mathcal{F}^{(a),\text{max}}_{2j}=-\frac{2(j+\frac 1 2)(2j+1)!!\zeta(2j+1)}{(2\pi)^j h^{2j+2}}\sqrt{\frac{(2j+2)^{2j+2}}{(2j+3)^{2j+3}}}
\end{equation}
is obtained at $a=\frac {h}{\sqrt{2j+2}}$.
The force in both cases is attractive.
The results for $\mathcal{F}^{(h)}$ are similar to those of $\mathcal{F}^{(a)}$ because of the symmetry between $a$ and $h$.

In Table \ref{helixtable1},
the Casimir energy and forces in the two directions for $D=2,3,4,5$ are listed.
\begin{table}[htp!]
\renewcommand{\arraystretch}{2.5}
\tbl{The massless helix Casimir energy and forces of a scalar field for $D=2,3,4,5$.}
{\begin{tabular}{@{}c|c|c|c@{}}
	\hline\hline
	$D$ & $\mathcal{E}_D^\text{reg.}$ & $\mathcal{F}^{(a)}_D$ & $\mathcal{F}^{(h)}_D$\\
	\hline
	2\hphantom{00}&\hphantom{0}$-\frac{\zeta(3)}{2\pi }\frac{1}{(a^2+h^2)^{\frac 3 2}}$\hphantom{00}&\hphantom{0}$-\frac{3 \zeta(3)}{2\pi }\frac{a}{(a^2+h^2)^{-\frac 5 2}}$\hphantom{00}&\hphantom{0}$-\frac{3 \zeta(3)}{2\pi }\frac{h}{(a^2+h^2)^{-\frac 5 2}}$\\
	\hline
	3\hphantom{00}&\hphantom{0}$-\frac{\pi ^2}{90 }\frac{1}{(a^2+h^2)^2}$\hphantom{00}&\hphantom{0}$-\frac{2\pi ^2}{45 }\frac{a}{(a^2+h^2)^3}$\hphantom{00}&\hphantom{0}$-\frac{2\pi ^2}{45 }\frac{h}{(a^2+h^2)^3}$\\
	\hline
	4\hphantom{00}&\hphantom{0}$-\frac{3\zeta(5)}{4\pi ^2 }\frac{1}{(a^2+h^2)^{\frac 5 2}}$\hphantom{00}&\hphantom{0}$-\frac{15 \zeta(5)}{4\pi ^2 }\frac{a}{(a^2+h^2)^{-\frac 7 2}}$\hphantom{00}&\hphantom{0}$-\frac{15 \zeta(5)}{4\pi ^2 }\frac{h}{(a^2+h^2)^{-\frac 7 2}}$\\
	\hline
	5\hphantom{00}&\hphantom{0}$-\frac{2\pi ^3}{945}\frac{1}{(a^2+h^2)^3}$\hphantom{00}&\hphantom{0}$-\frac{4 \pi ^3}{315}\frac{a}{(a^2+h^2)^4}$\hphantom{00}&\hphantom{0}$-\frac{4 \pi ^3}{315}\frac{h}{(a^2+h^2)^4}$\\
	\hline
\end{tabular}}
\label{helixtable1}
\end{table}

\begin{figure}[htp!]
\centering
\includegraphics[width=0.6\textwidth]{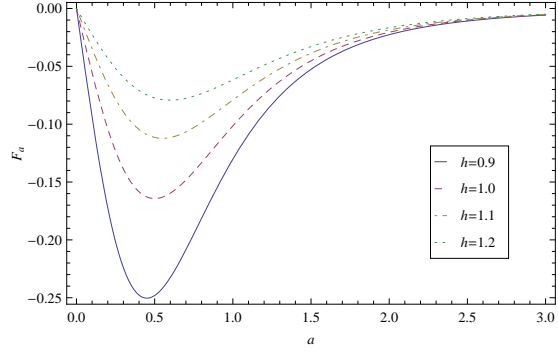}
\caption{The Casimir force on the $x$ direction \textit{vs.}$a$ in
$D=3$ dimension for different $h$. The Casimir force decreases with
$h$ increasing and the maximum value of the force magnitude
$\frac{2\pi^2}{45h^5}\sqrt{\frac{5^5}{6^6}}$ appears at
$a=\frac{h}{\sqrt{5}}$.}
\label{helixfig1}
\end{figure}

Fig. \ref{helixfig1} is the illustration\cite{ZHAI2011a} of the behavior of the Casimir force on
$x$ direction in $D=3$ dimension. The curves from the bottom to top
correspond to $h=0.9,1.0,1.1,1.2$ respectively. It is clearly seen that the Casimir force decreases with $h$ increasing and the maximum value of the force magnitude
$\frac{2\pi^2}{45h^5}\sqrt{\frac{5^5}{6^6}}$ appears at
$a=\frac{h}{\sqrt{5}}$.

\begin{figure}[htp!]
\centering
\includegraphics[width=0.6\textwidth]{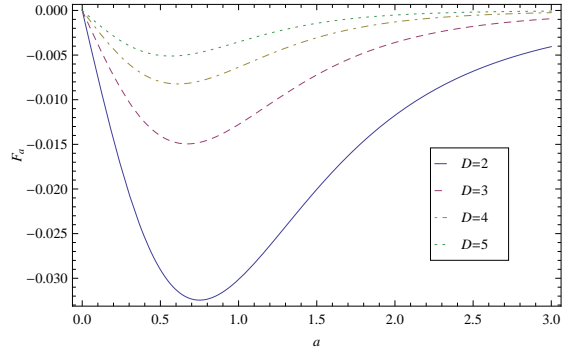}
\caption{The Casimir force on the $x$ direction \textit{vs.}$a$ in
different dimensions. It is set $h=1.5$. It is clearly seen that
the Casimir force decreases with $D$ increasing, and the value of
$a$ where the maximum value of the force is achieved also gets
smaller with $D$ increasing.}
\label{helixfig2}
\end{figure}

Fig. \ref{helixfig2} is the illustration\cite{ZHAI2011a} of the behavior of the Casimir force
on $x$ direction in different dimensions.
The curves from the bottom to top correspond to $D=2,3,4,5$ respectively.
It is set $h=1.5$ in this figure.
It is clearly seen that the Casimir force decreases with $D$ increasing,
and the value of $a$ where the maximum value of the force is achieved also gets smaller with $D$ increasing.

\subsubsection{Massive scalar field}
The massive case is slightly different in the eigenmodes with the mass $\mu$
\begin{eqnarray}\label{frequency}
    \omega_n^2 &=& k_{T}^2 + k_x^2 + \left( -\frac{2\pi n}{h}+\frac{k_x}{h}a \right)^2+\mu^2\nonumber\\
     &=& k_{T}^2 + k_z^2 + \left( \frac{2\pi n}{a}+\frac{k_z}{a}h
    \right)^2+\mu^2 \,.
\end{eqnarray}
\noindent where $k_x$ and $k_z$ satisfy eq.\eqref{kxkzf}.
The Casimir energy density of the massive scalar field in the ($D+1$)-dimensional spacetime is thus given by
\begin{eqnarray}
	\mathcal{E}^{\mu}_{D}=\frac1{2a} \int \frac{d^{D-1}k}{(2\pi)^{D-1}} \sum_{n=-\infty}^{\infty} \sqrt{k_T^2 + k_z^2 + \left( \frac{2\pi n}{a}+\frac{k_z}{a}h \right)^2 +\mu^2 }
	\label{tot energy-m}
\end{eqnarray}

To regularize eq.\eqref{tot energy-m}, the functional relation
\begin{equation}
	\begin{split}
	&\sum_{n=-\infty}^{\infty} \left (bn^2+\mu^2\right )^{-s}\\
	=&\frac{\sqrt{\pi}}{\sqrt{b}}\frac{\Gamma\left (s-\frac 12\right )}{\Gamma(s)}\mu^{1-2s}+\frac{\pi^s}{\sqrt{b}}\frac{2}{\Gamma(s)}\sum_{n=-\infty}^{\infty}{'}\mu^{\frac 12-s}\left (\frac{n}{\sqrt{b}}\right )^{s-\frac 12}K_{\frac 12-s}\left( 2\pi \mu \frac{n}{\sqrt{b}}\right )
	\end{split}
\end{equation}
is used, where the prime means that the term $n=0$ has to be excluded.
After tedious deduction, one arrives
\begin{equation}
	\mathcal{E}^{\mu,\text{reg.}}_D=-\frac{\mu^{D+1}\Gamma\left (-\frac{D+1}{2}\right)}{2^{D+2}\pi^{\frac{D+1}{2}}}-2\left(\frac{\mu}{2\pi\sqrt{a^2+h^2}}\right)^{\frac{D+1}{2}}\sum_{n=1}^{\infty}n^{-\frac{D+1}{2}}K_{\frac{D+1}{2}}\left( n\mu \sqrt{a^2+h^2}\right ).
\end{equation}
Utilize the asymptotic behavior $K_{\nu}(z)\rightarrow\frac{2^{\nu-1}\Gamma(\nu)}{z^{\nu}}$ when $z\rightarrow 0$ for $\nu>0$,
it is not difficult to recover the result of the massless case when $\mu\rightarrow 0$.

Using $K_{\nu}^{\prime}(z)=\frac{\nu}{z}K_{\nu}(z)-K_{\nu+1}(z)$
where $K_{\nu}^{\prime}(z)=dK_{\nu}(z)/dz$,
one has the Casimir force
\begin{equation}
	\mathcal{F}^{(a),\mu}=-\frac{2\mu a\left ( (\mu a)^2+(\mu h)^2\right)^{\frac{D+1}{4}}}{\left ( 2\pi\right )^{\frac{D+1}{2}}\left(a^2+h^2\right )^{\frac{D+2}{2}}}\sum_{n=1}^{\infty}n^{-\frac{D-1}{2}}K_{\frac{D+3}{2}}\left( n\mu \sqrt{a^2+h^2}\right ).
\end{equation}

\begin{figure}[htp!]
\centering
\includegraphics[width=0.6\textwidth]{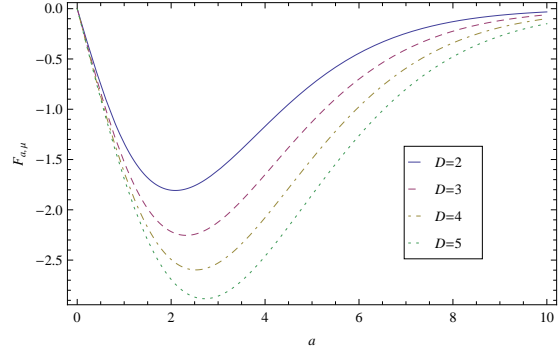}
\caption{The Casimir force on the $x$ direction \textit{vs.}$a$
in different dimensions for a massive scalar field.
It is taken $h=1,\mu=1$ and $D=2,3,4,5$ respectively.
It is clearly seen that the Casimir force increases with $D$ increasing,
and the maximum value of the force moves to larger $a$ as as $D$ increasing,
which is a feature that opposite to massless case.}
\label{helixfig3}
\end{figure}

Numerical analysis of the behavior of the Casimir force on $x$ direction
as a function of $a$ for different $h$ and $D$ can be found in Ref. \refcite{ZHAI2011a}.
The Casimir force is found still attractive
and it has a maximum value similarly to massless case.
For given values of $D$ and $\mu$,
the behavior of the force for different $h$ is similar to that in massless case.
But for given values of $h$ and $\mu$,
the behavior of the force for different $D$ is opposite to that in massless case.
The force increases with $D$ increasing
and the position of the maximum value move to larger $a$ as $D$ increasing.
Fig. \ref{helixfig3} shows the force as a function of $a$ for $\mu=1,h=1$ and $D=2,3,4,5$ respectively
and it is easy to find the difference between Fig. \ref{helixfig3} and Fig. \ref{helixfig2}.

The rate of massive and massless cases is given as follows
to study the precise way the Casimir force varies as the mass changes.
\begin{equation}
\frac{\mathcal{F}^{(a),\mu}}{\mathcal{F}^{(a),0}}=\frac{\left ( (\mu a)^2+(\mu h)^2\right
)^{\frac{D+3}{4}}}{(D+1)2^{\frac {D-1}{2}}\Gamma\left ( \frac
{D+1}{2}\right
)\zeta(D+1)}\sum_{n=1}^{\infty}n^{-\frac{D-1}{2}}K_{\frac{D+3}{2}}\left
( n\mu \sqrt{a^2+h^2}\right ).
\label{helixforceratio}
\end{equation}
In the case of odd-dimensional space, eq.\eqref{helixforceratio} can be reduced to
\begin{equation}
\frac{\mathcal{F}^{(a),\mu}}{\mathcal{F}^{(a),0}}=\frac{2(2j+1)!!\left ( (\mu a)^2+(\mu
h)^2\right )^{\frac{j+2}{2}}}{\left (2\pi\right
)^{2j+2}|B_{2j+2}|}\sum_{n=1}^{\infty}n^{-j}K_{j+2}\left ( n\mu
\sqrt{a^2+h^2}\right ),
\end{equation}
where $j=1,2,...$.
Obviously, the ratio tends to 1 when $\mu\rightarrow 0$
and it tends to zero when $\mu\rightarrow \infty$.

\begin{figure}[htp!]
\centering
\includegraphics[width=0.6\textwidth]{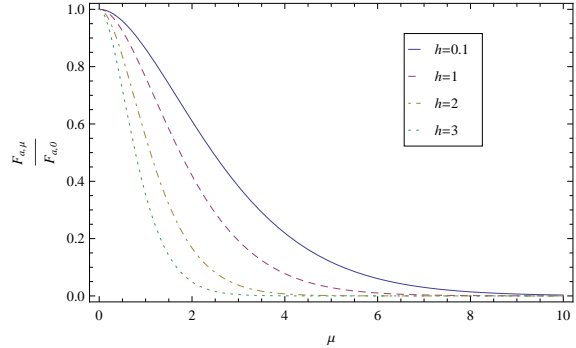}
\caption{the ratio of the Casimir force in massive case to that in massless case varying with the mass for different $h$ in $D=3$ dimension. The curves correspond to $a=1$ and $h=0.1,1,2,3$ respectively.}
\label{helixfig4}
\end{figure}

\begin{figure}[htp!]
\centering
\includegraphics[width=0.6\textwidth]{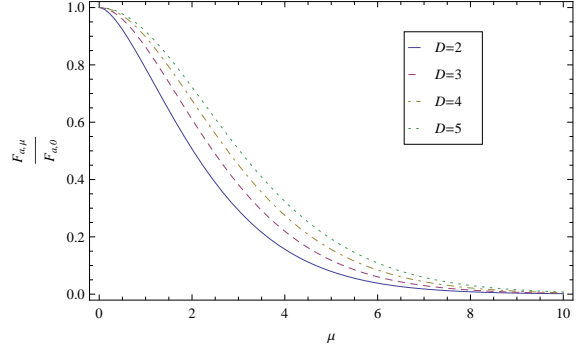}
\caption{the ratio of the Casimir force in massive case to that in massless case varying with the mass for different dimensions. We take $a=1$ and $h=0.1$. }
\label{helixfig5}
\end{figure}

Fig. \ref{helixfig4} is the illustration of the ratio of the Casimir force in
massive case to that in massless case varying with the mass in $D=3$
dimension. The curves correspond to $a=1$ and $h=0.1,1,2,3$
respectively.
Fig. \ref{helixfig5} is the illustration of the ratio of the Casimir
force in massive case to that in massless case varying with the mass
for different dimensions. The curves correspond to $a=1, h=0.1$ and
$D=2,3,4,5$ respectively. It is clearly seen from the two figures
that the Casimir force decreases with $\mu$ increasing, and it
approaches zero when $\mu$ tends to infinity. The plots also tell us
the Casimir force for a massive field decreases with $h$ increasing
but it increases with $D$ increasing. For the latter, the behavior
of the Casimir force in massive case is different from that of
massless case.

\subsection{Fermion Casimir Effect }
\subsubsection{The Vacuum Energy Density for a Fermion Field }
For the fermion field,
first a topological space $X$ as follows
\begin{equation}
X=\bigcup_{\mathbf{u}\in \mathrm{\Lambda}''}\{C_0+\mathbf{u}\}
\end{equation}
is considered in $\mathcal{M}^{D+1}$ with the induced topology and define an
equivalence relation $\sim$ on $X$ by
\begin{equation}
(x^1,x^2)\sim(x^1-2a, x^2+2h),
\end{equation}
then $X/\sim$ with the quotient topology is homomorphic to helix
topology. Here, $\mathrm{\Lambda}''$ and unit cylinder-cell $C_0$\cite{Feng2010,ZHAI2011a} are
\begin{equation}\label{sub2f}
  \mathrm{ \Lambda}'' = \left\{ ~ n(\mathbf{e}_2 - \mathbf{e}_1) ~|~ n \in \mathbb{Z} ~\right\} \,.
\end{equation}
and
\begin{eqnarray}
   C_0 &=& \bigg\{ \sum_{i=0}^{D}x^i \mathbf{e}_i ~|~ 0\leq x^1 < a,
 -h\leq x^2 < 0 ,\nonumber\\
 &-&\infty <x^0<\infty, -\frac{L}{2} \leq x^T\leq \frac{L}{2}\bigg\} \,,\label{cellf}
\end{eqnarray}
where $T = 3,\cdots, D$.
Then the anti-helix conditions imposed on a field $\psi$,
\begin{equation}
\psi(t, x^1+a,x^2,x^T)=-\psi(t,x^1,x^2+h,x^T)
\label{helixboundaryfermion}
\end{equation}
is considered,
where the field returns to the same value after traveling distances
$2a$ at the $x^1$-direction and $2h$ at the $x^2$-direction.
It is notable that a spinor wave function is anti-helix
and takes its initial value after traveling distances $2a$ and $2h$ respectively.
In other words, the anti-helix conditions are imposed on the field,
which returns to the same field value $\psi(t,
x^1+2a,x^2,x^T)=\psi(t,x^1,x^2+2h,x^T)$ only after two round trips.
Therefore, the BC \eqref{helixboundaryfermion} can be induced by $X/\sim$
with the quotient topology.

A spin-1/2 field $\psi(t, x^\alpha, x^T)$ defined
in the ($D+1$)-dimensional flat space-time satisfies the Dirac equation:
\begin{equation}\label{eomf}
    i\gamma^{\mu}\partial_{\mu}\psi - m_0\psi = 0 \,,
\end{equation}
where $\alpha=1,2; T=3,\cdots, D$; $\mu=(t,\alpha,T)$
and $m_0$ is the mass of the Dirac field.
$\gamma^{\mu}$ are $N\times N$ Dirac matrices with $N=2^{[(D+1)/2]}$
where the square brackets mean the integer part of the enclosed expression.
It is assumed in the following that these
matrices are given in the chiral representation:
\begin{equation}
\gamma^0=\Bigg(\begin{array}{clrr}
      1 &\hspace{0.2cm} 0  \\       0  & -1  \end{array}\Bigg),\gamma^k=\Bigg(\begin{array}{clrr}
      0 & \sigma_k  \\       -\sigma_k^{+}  & 0  \end{array}\Bigg), k=1,2,\cdots, D
\end{equation}
with the relation
$\sigma_{\mu}\sigma_{\nu}^{+}+\sigma_{\nu}\sigma_{\mu}^{+}=2\delta_{\mu\nu}$.
Under the BC eq.\eqref{helixboundaryfermion}, the solutions of the field can be
presented as
\begin{equation}
\psi^{(+)}=\mathcal{N}^{(+)}e^{-i\omega t}\Bigg(\begin{array}{clrr}
      e^{i(k_x x+k_z z+k_T x^T)}\varphi_{(\alpha)}  \\    -i\mbox{\boldmath$\sigma$}^{+}\cdot\mbox{\boldmath$\nabla$} e^{i(k_x x+k_z z+k_T x^T)}\varphi_{(\alpha)} /(\omega+m_0)   \end{array}\Bigg),
      \label{positivesolution}
\end{equation}
and
\begin{equation}
\psi^{(-)}=\mathcal{N}^{(-)}e^{i\omega t}\Bigg(\begin{array}{clrr}
       i\mbox{\boldmath$\sigma$}\cdot \mbox{\boldmath$\nabla$}e^{i(k_x x+k_z z+k_T x^T)}\chi_{(\alpha)} /(\omega+m_0) \\  e^{i(k_x x+k_z z+k_T x^T)}\chi_{(\alpha)}     \end{array}\Bigg),
       \label{negativesolution}
\end{equation}
where $\mbox{\boldmath$\sigma$}=(\sigma_1,\cdots, \sigma_D), x^1=x,
x^2=z$ and $\mathcal{N}^{(\pm)}$ is a normalization factor, and
$\varphi_{(\alpha)}, \chi_{(\alpha)}$ are one-column constant matrices
having $2^{[(D+1)/2]}-1$ rows with the element
$\delta_{\alpha\beta}, \alpha,\beta=1,\cdots, 2^{[(D+1)/2]}-1$.

From eqs.\eqref{eomf}-\eqref{negativesolution}, one has
\begin{equation}
	\begin{split}
		\omega_n^2 =& k_{T}^2 + k_x^2 + \left( -\frac{2\pi (n+\frac 12)}{h}+\frac{k_x}{h}a \right)^2+m_0^2\\
		=& k_{T}^2 + k_z^2 + \left( \frac{2\pi (n+\frac 12)}{a}+\frac{k_z}{a}h  \right)^2+m_0^2,
	\end{split}
    \label{energy}
\end{equation}
with $k_x$ and $k_z$ satisfying
\begin{equation}\label{kxkz}
    a k_x - hk_z = 2\left (n+\frac 12\right )\pi\,, (n=0,\pm1,\pm2,\cdots).
\end{equation}
The energy density of the field in
($D+1$)-dimensional space-time is thus given by
\begin{equation}
	\mathcal{E}_{D}=-\frac{N}{2 a} \int \frac{d^{D-1}k}{(2\pi)^{D-1}}\sum_{n=-\infty}^{\infty} \sqrt{k_T^2 + k_z^2 + \left( \frac{2\pi (n+\frac 12)}{a}+\frac{k_z}{a}h \right)^2 +m_0^2 },
 	\label{tot energy}
\end{equation}
where it is also assumed $a\neq 0$ without losing generalities.

Eq.\eqref{tot energy} can be rewritten as
\begin{equation}
	\begin{split}
		\mathcal{E}_D=&- \frac{N}{2 a \sqrt{\gamma}}  \int \frac{d^{D-1}u}{(2\pi)^{D-1}} \sum_{n=-\infty}^{\infty} \sqrt{u^2 + \left( \frac{2\pi (n+\frac 12)}{a \sqrt{\gamma}}\right)^2+m_0^2  }\\
		=&\frac{2^{[(D+1)/2]-(D+1)}\Gamma\left ( -\frac D 2 \right)}{\pi ^{\frac D 2}a\sqrt{\gamma}}\sum_{n=-\infty}^{\infty}\left[\left (\frac{2 \pi (n+\frac 12)}{a \sqrt{\gamma}} \right)^2+m_0^2\right ]^{\frac D2},
	\end{split}
	\label{re-tot energy}
\end{equation}
with $\gamma \equiv 1+ \frac {h^2}{a^2}$.

It is seen from eq.\eqref{re-tot energy} that the expression for the vacuum energy
in the case of helix BCs can be obtained from the
corresponding expression in the case of standard BC
$\psi (t,x^1+a,x^2,x^T)=-\psi (t,x^1,x^2,x^T)$ by making the change
$a \rightarrow a\sqrt{\gamma }=\sqrt{a^2+h^2}$. The topological
fermionic Casimir effect in toroidally compactified space-times has
been recently investigated in Ref. \refcite{Bellucci2009a} for non-helix
BCs including general phases. In the limiting case
$h=0$, our result of eq.\eqref{re-tot energy} is a special case of general formulas
from Ref. \refcite{Bellucci2009a}.

\subsubsection{The Case of Massless Field}
For a massless Dirac field, that is, in the case of $m_0=0$, the
energy density in eq.\eqref{re-tot energy} is reduced to
\begin{equation}\label{fienergy}
	\mathcal{E}_{D}^{0}=\frac {2^{[(D+1)/2]} \pi^{\frac D 2}}{a^{D+1}\gamma^{\frac
{D+1}{2}}}\Gamma \left (-\frac D 2 \right )\zeta(-D,\frac 12).
\end{equation}

\noindent where $\zeta(-D,\frac 12)$ is the Hurwitz-Riemann $\zeta$
function. Using the relation
\begin{equation}
\zeta(s,\frac 12)=(2^s-1)\zeta(s),
\end{equation}
and the reflection relation eq.\eqref{RiemannZetaReflect},
the energy density can be regularized to be
\begin{equation}\label{fienergy1}
	\mathcal{E}_D^{0,\text{reg.}}=2^{[(D+1)/2]} \left (2^{-D}-1\right )\frac {\Gamma \left
(\frac {D+1} 2 \right )\zeta(D+1)}{\pi^{\frac {D+1}
2}(a^2+h^2)^{\frac{D+1}2}}.
\end{equation}
The Casimir force on the $x$ direction is
\begin{equation}
	\begin{split}
		\mathcal{F}_D^{0,(a)}=&-\frac {\partial\mathcal{E}_D^{0,\text{reg.}}}{\partial a}\\
		=&2^{[(D+1)/2]} \left (2^{-D}-1\right )\frac {(D+1)\Gamma \left (\frac {D+1} 2\right)\zeta(D+1)}{\pi^{\frac {D+1} 2}} \frac{a}{(a^2+h^2)^{\frac{D+3}2}}.
	\end{split}
\end{equation}
It is obvious that the energy density is negative and the
force is attractive. Furthermore, the force has a maximum value
\begin{equation}
	\begin{split}
		\mathcal{F}_D^{0,(a),\text{max}}=&2^{[(D+1)/2]} \left (2^{-D}-1\right )\frac{(D+1)\Gamma \left (\frac {D+1} 2 \right )\zeta(D+1)}{\pi^{\frac{D+1}2}h^{D+2}}\sqrt{\frac{(D+2)^{D+2}}{(D+3)^{D+3}}}
	\end{split}
\end{equation}
at $a=\frac{h}{\sqrt{D+2}}$. The results for $\mathcal{F}_D^{0,(h)}$ are similar
to those of $\mathcal{F}_D^{0,(a)}$ because of the symmetry between $a$ and $h$.

\subsubsection{The Case of Massive Field}
For a massive Dirac field, to regularize the series in eq.\eqref{re-tot energy}
the Chowla-Selberg formula is used directly\cite{Elizalde1998}
\begin{equation}
	\begin{split}
		\sum_{n=-\infty}^{\infty}\left [\frac 1 2 a(n+c)^2+b\right ]^{-s}=&\frac{(2\pi)^{\frac 1 2}b^{\frac 1 2-s}}{\sqrt{a}}\frac{\Gamma\left(s-\frac 1 2\right)}{\Gamma(s)}+\frac{2^{\frac s 2 +\frac 1 4+2}\pi^sb^{-\frac s 2+\frac 14}}{\sqrt{a}\Gamma(s)}\\
		&\times\sum_{n=1}^{\infty}\cos(2\pi nc)\left(\frac{n^2}{a}\right)^{\frac s 2-\frac 1 4}K_{\frac 1 2-s}\left(2\pi n\sqrt{\frac{2b}{a}}\right).
	\end{split}
	\label{CS3}
\end{equation}
Note that in the renormalization procedure, the vacuum energy in a flat
space-time with trivial topology should be renormalized to zero,
that is, in the expression for the renormalized vacuum energy the
term corresponding to the first term in the right hand side of eq.\eqref{CS3} should be omitted.
Thus, the Casmir energy has the expression as follows
\begin{equation}
	\begin{split}
		\mathcal{E}_D^{m_0,\text{reg.}}=&2^{[(D+3)/2]}\left(\frac{m_0}{2\pi\sqrt{a^2+h^2}}\right )^{\frac{D+1}{2}}\\
		&\times\sum_{n=1}^{\infty}\cos(\pi n)n^{-\frac{D+1}{2}}K_{\frac{D+1}{2}}\left ( nm_0\sqrt{a^2+h^2}\right).
	\end{split}
\end{equation}
One can also find that the energy recovers the massless result by use of the asymptotic behavior of $K_\nu(z)$.

And similarly the Casimir force
\begin{equation}
	\begin{split}
		\mathcal{F}^{(a),m_0}=&\frac{2^{[(D+3)/2]} m_0 a\left ( (m_0 a)^2+(m_0h)^2\right )^{\frac{D+1}{4}}}{(2\pi)^{\frac{D+1}{2}}\left (a^2+h^2\right )^{\frac {D+2}{2}}}\\
		&\times\sum_{n=1}^{\infty}\cos(\pi n)n^{-\frac{D-1}{2}}K_{\frac{D+3}{2}}\left ( n m_0\sqrt{a^2+h^2}\right ).
	\end{split}
\end{equation}
The influence of mass on the Casimir force is similar to that of the case of scalar field.
Numerical analysis can be found in Ref. \refcite{Zhai2011}.

We have reviewed in this section the scalar and fermionic quantum spring in $(D+1)$-dimensional spacetime using the zeta function techniques.
The Casimir force of both the scalar and spinor field is attractive,
and has a maximum of magnitude.
The influence of mass of the field on the Casimir effect has also been reviewed for both fields,
and the precise way of the Casimir force changing with the mass has given.

\section{Summary and Outlook}
\label{summary}
The quantum property of the reality,
may be one of the most mystic, revolutionary but also fascinating concepts
that physics theory has ever been brought to us.
The Casimir effect provides a possibility to have a direct and perhaps macroscopic access
to the insight of this reality,
which makes the topic of this effect still full of vigor and vitality
after its discovery more than 60 years ago.
Study of the Casimir effect in rectangular boxes,
one of the typical configuration of this topic,
captures a lot of features of the effect.

In this article, we have reviewed several researches
related to the Casimir effect in rectangular boxes.
The frequently used regularization methods,
the zeta function and Abel-Plana formula techniques,
are proven identifiable,
which gives our freedom to choose any regularization method at convenient.
The equivalence of these two approaches may be extended to
other regularization methods and other configurations.
With the powerful zeta function technique,
we have summarized the attractive and repulsive nature
of the Casimier effect in rectangular boxes with various settings,
and in addition, reviewed the rectangular Casimir piston.
Unlike the boxes, Casimir forces on the piston are always attractive no matter how
the ratios of edges change.
These researches have probed into the very nature
of the quantum field on vacuum state.
And furthermore, the study of the Casimir effect of quantum field on
non-vacuum states, the equilibrium state characterized by a finite temperature
has also been presented.
The thermal Casimir effect, compared to the zero point energy,
is a more practical and feasible subject,
since an ensemble and distribution of excited states is a typical situation
and almost all experiments are done under some temperature.
The configuration of hypercube is only a simplified example
to illustrate that both the zero temperature and temperature-dependent parts
of the free energy need to be regularized.
Temperature corrections on the effect for the configuration suggested in
Sect. \ref{pmbox} and \ref{piston} are worth looking into.
Finally, we have reviewed the Casimir effect arising from the non-trivial topology of the space,
the quantum spring for both scalar and fermion fields,
in which it is found that the forces are always attractive
and have a maximum of magnitude.
In practice, the study quantum spring may be applied to microelectromechanical system (MEMS).

Fragmental may be these researches, we have seen different aspects of the nature
of the Casimir effect.
These deepen understandings of the quantum nature,
pieces of which may be united in future,
have found their applications in various areas of research,
including both fundamental physics and applied science.
Extensive and detailed study of the subject suggests that
its infancy is over and it is progressing towards its mature.
We hope our review will serve as a collection of a part of
the resource for its future development.
\section*{Acknowledgments}
This work is partially supported by National Science Foundation of China grant Nos.~11105091 and~11047138,
``Chen Guang" project supported by both Shanghai Municipal Education Commission and Shanghai Education Development Foundation Grant No. 12CG51,
Shanghai Natural Science Foundation, China grant No.~10ZR1422000,
Key Project of Chinese Ministry of Education grant, No.~211059,
Shanghai Special Education Foundation, No.~ssd10004.
and Program of Shanghai Normal University (DXL124).

%\begin{thebibliography}{000} %for 3 digits
%\begin{thebibliography}{00}  %for 2 digits
\bibliography{refs}
\bibliographystyle{ws-ijmpa}
\end{document}